# Casimir-like effect driven self-assembly of graphene on molten metals

Kristýna Bukvišová[1,2], Radek Kalousek[1,3], Marek Patočka[3], Jakub Zlámal[1,3], Jakub Planer[1], Vojtěch Mahel[2,3], Daniel Citterberg[3], Libor Novák[2], Tomáš Šikola[1,3], Suneel Kodambaka[4], Miroslav Kolíbal[1,3*]

[1]CEITEC BUT, Brno University of Technology, Purkyňova 123, 612 00 Brno, Czech Republic

[2]Thermo Fisher Scientific, Vlastimila Pecha 12, 627 00 Brno, Czech Republic

[3]Institute of Physical Engineering, Brno University of Technology, Technická 2, 616 69 Brno, Czech Republic

[4]Department of Materials Science and Engineering, Virginia Polytechnic Institute and State University, Blacksburg, Virginia 24061, United States

*kolibal.m@fme.vutbr.cz

**Casimir effect, explained by Hendrik Casimir in 1948, is a macroscopic manifestation of quantum electrodynamics. Symmetry breaking due to space confinement of vacuum fluctuations in between two planar mirrors results in an attractive force acting between the two mirrors. Here, we show that spontaneous self-assembly of two-dimensional (2D) layered materials grown on molten metals (rheotaxy) is driven by the mechanical forces exerted by the liquid on the graphene domain and that these forces are subject to a Casimir-like effect. We present in situ environmental and ultrahigh vacuum scanning electron microscopy observations of chemical vapor deposition of graphene on both solid and molten gold and copper, which reveal that the self-assembly occurs via translational and rotational motions of graphene domains during growth on molten metals. Using high-temperature (~ 1300 K) atomic force microscopy measurements of graphene/molten-metal interfaces, coupled with density functional theory and continuum modelling of 2D layers floating on a liquid surface, we attribute the observed phenomena to Casimir-like effect of surface undulations and give further guidance for achieving seamless stitching of 2D layer domains into large scale monolayers.**

Since the discovery of graphene, considerable efforts have been aimed at the production of large-area graphene sheets. Among the several different synthesis approaches proposed to-date, chemical vapour deposition (CVD) has been the most promising method for obtaining high-quality large-area layers of graphene. Using single-crystalline substrates and by the optimal choice of growth parameters, large-domain and single-crystalline graphene layers have been obtained.(*1–5*). CVD onto polycrystalline foils and amorphous substrates, however, typically yield polydomain graphene, i.e. with multiple rotational domains separated by boundaries; such graphene layers are undesirable for most applications. Recently, rheotaxy -- growth of crystalline thin films on molten surfaces (*6–10*) -- has been used to produce graphene layers with highly ordered domains *via* CVD on molten Cu (*11, 12*). The absence of rigid substrate has been proposed (*13*) to promote strain-free (*14*) and seamless assembly of many grains that originated from numerous nucleation events during growth. As such, rheotaxy holds great promise even beyond the realm of graphene (*15–17*). The successful demonstration of large-area graphene growth via rheotaxy has motivated *in situ* studies, which have suggested that short-range electrostatic interactions and long-range capillary forces may be the factors contributing to self-assembly (*18–20*), however, the exact mechanisms are not known. Self-assembly has been observed previously in different nanoscale systems (*21*), being driven by e.g. strain relief in 'Stranski-Krastanov' growth (*22*) or electron density standing waves (*23*). At the mesocale, the long-range forces are required to arrange e.g. domains of 2D materials into regular patterns (*24*). Here, we focus on fundamental understanding of the dynamics of graphene domain self-assembly on molten metals, with the aim of providing a general description of the self-assembly processes on liquid substrates.

In order to study the dynamics of self-assembly, we have employed a suite of *in situ* characterization tools, scanning electron microscopy (SEM) for monitoring the growth of graphene on molten metals and for the *first time* high-temperature atomic force microscopy (HT-AFM) to determine the graphene/melt interfacial interactions, and modelling to gain insights into the observed phenomena. In the following sections, we first showcase our *in situ* monitoring capabilities for the direct observation of graphene growth before presenting our results and analyses of the phenomenon of self-assembly.

We carried out graphene growth and monitoring experiments *in situ* in two different microscopes using ethylene as the carbon source. One set of experiments involved the growth of graphene layers on Au using a high-pressure (up to 150 Pa) MicroReactor (*25, 26*) and the other on Cu in an ultra-high vacuum (UHV, base pressure ~$10^{-7}$ Pa) environment at temperatures (1073 K < $T$ < 1390 K) above and below their melting points ($T_{m, Cu}$ = 1357 K and $T_{m, Au}$ = 1336 K), where the metals are either liquid or solid, respectively. The MicroReactor design (*25*) (see fig. S1 in Supporting Material) enables simultaneous heating of the metal samples to high temperatures ($T > T_m$) and dosing reactive gases at high pressures during the operation of SEM, which facilitates the observation of graphene nucleation and growth on solid and molten Au. The UHV experiments were carried out in a custom-designed SEM, in which we melted copper supported on a platinum wire that was used as a resistive heater (sample design shown in fig. S1), with the base pressure ~ $10^{-7}$ Pa to ensure the cleanliness of the reactive molten surface. We chose copper because it is the most common metal catalyst for graphene CVD growth and was also used for graphene rheotaxy (*18, 20, 27*). Although gold is not a commonly used catalyst for graphene growth, we chose gold due to its very low vapor pressure (~$10^{-3}$ Pa at $T$ = 1200 K, (*28*)) and, more importantly, due to its inertness toward oxidation. Using Raman spectroscopy (see fig. S1), we have confirmed that the layers deposited on both solid and liquid Au and Cu surfaces in our *in situ* experiments are graphene. Further, we have found that the kinetics of nucleation, growth, and shape evolution of graphene domains depend on both the catalyst (e.g., Au vs. Cu) and on its state (liquid vs. solid), which is discussed elsewhere.

Figure 1 shows a representative set of SEM images extracted from Movies S1 and S2 obtained during graphene growth on solid (Fig. 1A) and liquid (Fig. 1B) Au, respectively. In these images, darker and lighter grey contrast features are individual graphene domains and metal surfaces, respectively. On molten Au, we have observed spontaneous assembly of the graphene domains into regular arrays, as documented in Fig. 1B and Movie S2. (We have also observed similar phenomenon during graphene growth on molten Cu.) Although the domains are supported on a molten metal, the centres of mass of 'floating' domains remain in-place (see Fig. 1C) with inter-domain spacings $d$ maintaining a steady-state over time $t$, irrespective of the domain size (Fig. 1D). Intriguingly, the domains on the liquid surface (panel B) appear fuzzy in the *in situ* SEM images. Since the images are acquired by rastering the electron beam, any change in the instantaneous positions of the domains within each frame results in blurred contours. Movies S2 and S6 reveal that the domains wobble and rotate during growth. The motion of domains on the liquid metal is direct evidence that the domains are weakly bonded to molten metal surfaces during growth. We suggest (and justify below) that this domain motion leads to self-assembly.

Often, surface dynamical phenomena such as self-assembly, Ostwald ripening, etc. are attributed to capillary forces, which give rise to meniscus on liquid surfaces. For graphene floating on the molten metal, a liquid meniscus, if present, would be manifested by a rim around the individual domains in secondary electron images. We have observed the rim only for graphene on partially molten metal surfaces (see fig. S2, middle panel); however, on fully molten metals, the individual graphene domains and the surrounding liquid appear flat, homogenous and rim-free in SEM images for both gold and copper (fig. S1, middle panel). In order to further verify this conclusion, we used HT-AFM operated up to $T \sim 1380$ K. Fig. S3 shows an AFM topography map measured across the edge of a graphene domain floating on molten gold. Graphene appears to be floating *above* the liquid surface, as indicated by an increase in the surface height of the graphene domain relative to the substrate. Importantly, we do not observe any curvature across the domain. That is, the liquid does not exhibit a meniscus (resolvable with our HT-AFM) that could be associated with capillary forces, in agreement with the SEM images.

In order to identify the factors leading to self-assembly, we focus on the phenomenon of wobbling of graphene domains seen in SEM (Fig. 1B, Movie S2). Fig. 2A shows typical SEM images (extracted from Movie S3) of an individual graphene domain confined within a region bounded by neighbouring larger graphene domains obtained as a function of time *t* during graphene growth. The domain appears fuzzy during early stages of growth. We estimate the rates of the domain motion to be between $10^{-6}$ and $10^{-5}$ m/s on both Cu and Au (see SI for details). The domains appear sharper at later times, suggestive of reduced wobbling, which is directly correlated with the decreasing distance between the domains. During wobbling, the domains avoid merging and remain separate. As the domains grow larger, they inevitably attach rapidly either to the surrounding larger domain wall or coalesce with other adjacent domains (Fig. 1B, Fig. 2A). When attachment occurs, it disrupts the regular spacing of the domains, indicating a stochastic and metastable nature of the self-assembly (*24*). We suggest that the wobbling and the attachment of domains are consequences of repulsive and attractive forces, respectively, between the domains and their local environment. It seems that the domains floating on a liquid metal surface in close proximity to other domains reside within a potential well.

To better understand the dynamics of the domain motion, we have quantitatively determined the extent of wobbling from the *in situ* SEM image sequences (fig. S4). Fig. 2B is a plot of the wobbling amplitudes $A$ measured for growing domains of different sizes $r$ and separation distances ($R$-$r$). We find that $A$ decreases with decreasing free space ($R$-$r$). Importantly, $A$ scales only with the available free liquid space, i.e. ($R$-$r$), and is independent of $r$ (fig. S5), shape, composition of the liquid catalyst (i.e. Au or Cu) (see Fig. 2B), ethylene partial pressure (i.e. deposition flux, see fig. S6), experimental setup and imaging procedure (i. e. beam-induced effects, see fig. S7). Moreover, wobbling continues even during domain etching (fig. S6). We do not observe domain oscillations on pre-molten surface (fig. S8). Based on these data, we rule out the Brownian motion (fig. S5, (*29*)), gas flow or carbon adatom concentration gradients (fig. S6), and dynamic growth/dissolution of the domains at the atomic scale (fig. S9, (*30*)). We have observed a weak dependence on temperature: amplitudes increase with the increasing $T$ (within 1223 - 1373 K for Au and 1357 – 1473 K for Cu, see fig. S5). All the data obtained from domains of different sizes, in different environments, on Cu and Au collapse onto a single curve. A pronounced difference between the liquid Cu and Au substrates is that on Au, the wobbling ceases at a critical distance ($R$-$r$) of 47 ± 20 nm (see Fig. 2B and fig. S10); for Cu, the critical distance is larger, approximately 140 ± 60 nm, suggesting that the balance between attractive and repulsive forces between the domains is reached sooner on Cu. Vanishing of the wobbling is immediately followed by domain coalescence indicated by the dashed vertical lines in Fig. 2B, and the previously established assembly of the domains is always disrupted.

Our *in situ* observations of graphene rheotaxy reveal self-assembly (Fig. 1) of individual graphene domains. More interestingly, we find that the domains oscillate (Fig. 2A) and reorient (see Fig. 3A) during self-assembly. The observed phenomena can arise due to (i) electrostatic, (ii) van der Waals (vdW), (iii) capillary, and (iv) Casimir-like interactions due to nanoscale undulations of the molten metal surface. (See Modeling in Supporting Material for detailed description of the effects of all these forces.) Electrostatic dipoles, considered to be of key importance in the alignment of 2D materials on molten metals (*18*), develop at the graphene-metal interfaces due to differences in work functions of graphene and the metals(*31, 32*). The electrostatic charge distribution across the graphene domains, calculated using density functional theory (DFT) (See Modelling in Supporting Material, figs. S11—S13 and tables S1 and S2), is expected to result in repulsive interactions, and hence oscillations with amplitudes, that scale with the domain shape, size $r$, and ($R$-$r$). However, $A$ is experimentally found to be independent of $r$ and shape (fig. S5). The capillary forces have been speculated to induce a nanoscale 'Cheerios effect' (*18, 33*) for floating objects. But surface morphologies of graphene domains on molten metals measured using *in situ* HT-AFM and SEM (see figs. S1-S3) do not show menisci around the graphene domains, indicating that capillary forces, if present, are negligible. The vdW forces, resulting in attractive interactions between the graphene domains could counter the repulsive electrostatic force mentioned above. Our estimations (see Modelling in Supporting Material and fig. S14) yield vdW forces of the order of $10^{-19}$ N for domains of radii 200 nm, too weak to cause attraction of the domains at the distances ($10^1$-$10^2$ nm) observed in experiment. The rapid wobbling of domains could occur due to a variety of reasons (Brownian motion, gas or liquid flow, electron beam effect etc.), but its absence on a premolten surfaces (fig. S8) together with other experimental results (fig. S5-S9) makes us conclude that the wobbling is related to the motion of the liquid surface. Therefore, we rule out electrostatic, capillary, and/or vdW forces and propose forces arising from thermal fluctuations in the graphene-metal system as the primary mechanism(s) leading to the observed phenomena.

Nanoscale menisci can form either due to thermal fluctuations of the liquid surface (*34–36*) or electrostatic interactions between graphene and the molten metal (*37*), generating long-range interactions over large-areas (*38, 39*). The related deformation of the liquid surface exerts a force on the floating object, making it move and rotate (*38, 40*); the alignment of parallel long edges seen in our and others' experiments (*11, 18*) is a characteristic feature of such surface undulations. The existence of large-scale surface undulations on liquid surfaces has been confirmed by X-ray diffraction (*35*). Hence, we have set up an analytical model of undulations (manifested as surface waves with a wavenumber *k*) acting on a floating 2D domain surrounded by stationary domains (see Modeling in Supporting Material). An inherent feature of such a system in a confined space is a Casimir-like effect, where the limited number of waves with certain *k*-vectors pushes the domain in certain directions (due to the absence of counteracting waves on the other side of the domain) and, potentially, induces a torque. We assume that the waves appear only on a bare liquid, i.e. on the molten metal surfaces not covered by graphene. The waves are described by a solution of the 2D Helmholtz equation of the form, $\Delta u_k + k^2 u_k = 0$, where *u* is the displacement with the same Dirichlet boundary conditions of the zero displacement for the boundary and for the edges of the graphene domains of interest. These solutions form a set of eigenfunctions belonging to specific eigenvalues of *k*. The wave amplitudes of every eigenfunction are given by a resonant curve (similar to the theory of driven oscillations), see fig. S15. Different waves can appear at different sides of the domains resulting in both repulsive and attractive interactions between domains (see fig. S15). Since every wave carries certain momentum (and energy), the wave reflection is accompanied by a change in momentum manifested in forces $\vec{F}_i \propto \sum_k \left(\frac{\partial u_k}{\partial n_i}\right)^2$ acting at *i*-th point of the domain edge pushing perpendicularly against the domain edge. The translational and rotational motions of the domain are obtained by solving the equations of motion, $m\vec{a} = -b\vec{v} + \sum_i \vec{F}_i$ and $I\vec{\varepsilon} = -\beta\vec{\omega} + \sum_i \vec{r} \times \vec{F}_i$, respectively, along the entire domain edge. In the above equations, *m*, $\vec{a}$, and $\vec{v}$ are the mass, acceleration, and velocity of the graphene domain, respectively; *I*, $\vec{\varepsilon}$, and $\vec{\omega}$ are the moment of inertia, angular acceleration, and angular velocity, respectively with respect to the domain center of mass. The parameters *b* and *β* represent the linear and angular damping constants, respectively, and are a measure of the dynamic viscosity of the molten metal.

Our model correctly reproduces the experimentally observed behavior. We have simulated the wobbling of floating domains within confined spaces, mimicking the experimental geometries. Fig. 2C is a plot of calculated displacements of the domains as a function of the free (i.e. the region not covered by graphene) liquid surface, (*R-r*). The simulated profile is strikingly similar to the experimental data in Fig. 2B. As the domain grows larger in time, (*R-r*) decreases and the domain is dragged out of the near-equilibrium position due to rapidly increasing attractive forces between its edges and the nearby boundary. When attractive forces on one side of the domain prevail, the domain moves toward and eventually attaches to the adjacent boundary. As seen in the simulations (Movie S4) and confirmed in experiments (Fig. 1B, 2A), the domain geometry is disturbed due to additional torque forcing the domain to rotate during attachment. Both simulations (see Movie S4) and experiment (Figs. 1B and 2A and Movies S2 and S3) demonstrate that at the very final stage before the rapid coalescence, the domain experiences a large angular momentum, which disrupts its ideal position.

Movies S4 and S5 show the dynamics of hexagonal domains on molten surfaces with different magnitudes of damping forces. The damping is a measure of viscosity, which is for liquid Au at 1373 K $\approx$ 1.35× larger than that for liquid Cu at the same $T$ (*41, 42*). The black and orange curves in Fig. 2C, respectively, correspond to larger and smaller damping values. Despite different damping, the scaling profiles of domain displacements vs. ($R$-$r$) are nearly identical. More importantly, the domain expectedly attached sooner in case of smaller damping (smaller viscosity), as observed in experiments for the domains on liquid copper as compared to liquid gold (Fig. 2B).

When a domain is confined in an enclosed space, formed for example by other domains, it rotates to attain a preferred position, which for hexagonal shapes is with their sides parallel to each other. This is experimentally demonstrated in Fig. 3A, which shows rotation of an isolated hexagonal graphene domain floating on a liquid copper. Clearly, the domains move and rotate on the liquid surface (*13*) (see Movie S6 and also Movie S7); however, this motion is restricted by the repulsive interactions arising from the presence of other domains in vicinity. This phenomenon is also captured in our modeling. Fig. 3B shows a series of simulated images of a hexagonal domain floating and rotating on a liquid surface. The data are extracted from Movie S4 generated using our model. In agreement with the experimental observations, the domain rotates and reorients itself until it attains a metastable position, which is with its sides parallel to the domain boundary (Movie S6 and S7), ideal for seamless assembly.

Finally, we comment on the validity and significance of our model based on Casimir-like forces due to surface undulations. A direct observation of capillary undulations on molten metals is beyond time and lateral resolutions of both SEM and AFM used; nevertheless, our model predictions are in a qualitative agreement with the experimental observations of domain oscillations and rotations (see Figs. 2 and 3). A quantitative description of the observed phenomena requires detailed knowledge of the relation(s) between damping constants and viscosities of the molten material and amplitude and frequency of the surface undulations (*43*), which is beyond the scope of our work. Our simple model, which ignores electric dipole interactions, vdW and capillary forces, growth kinetics, etc., also nicely replicates more complex systems, e.g. coordinated behaviour of many domains as observed experimentally in Fig. 1A and Movies S1 and S2 (see simulated Movie S8). However, the influence of electric dipole interactions increases with the increasing domain size. Hence, we expect that the role of dipole-dipole interactions will be more important for graphene domains that are larger than those observed in this study(*11*, *18*). We note that our model is material independent, and hence is applicable to predict and explain self-assembly of 2D layers and even 3D crystals on surfaces of any liquids and weakly interacting materials capable of producing surface undulations.

**Conclusions**

We have investigated the phenomenon of self-assembly occurring during the chemical vapor deposition of graphene on molten metals such as copper and gold. Our *in situ* SEM observations reveal that individual graphene domains oscillate and rotate during growth as a means to arrange themselves in spatially-periodic arrays. Based on the time-resolved measurements of the domain dynamics as a function of their sizes and HT-AFM data of the graphene/molten-metal topography, we have ruled out electrostatic interactions and capillary forces and show that the Casimir-like effect of surface undulations can explain all our observations. Our data suggest that in order to achieve seamless stitching of domains, essential for large-area rheotaxy of single-crystalline sheets of 2D layers, among the possible molten materials that can be used as substrates, those with high viscosity are likely to yield better results.

**Acknowledgments.** We would like to thank Michal Dymáček for coding the image analysis software and Tomáš Spusta for rendering images shown in fig. S1 and S3. Access to the microscopes at Thermo Fisher Scientific facility in Brno and at the Nenovision company is greatly acknowledged.

**Funding**:

Quantum materials for applications in sustainable technologies (QM4ST) - project No. CZ.02.01.01/00/22_008/0004572 by OP JAK, call Excellent Research.

Brno University of Technology - specific research FSI-S-23-8336

Grant Agency of the Czech Republic (grant No. 23-07617S)

Ministry of education, Youth and Sports of the Czech Republic - CZ.02.2.69/0.0/0.0/18_053/0016962 (MEMOV II - International mobility of Brno University of Technology Researchers II)

Ministry of education, Youth and Sports of the Czech Republic - LM2023051

Air Force Office of Scientific Research (AFOSR, Dr. Ali Sayir) under Grant # FA9550-20-1-0184

National Science Foundation (NSF) for DMR Award 2245008 (old award ID 2211350)

U.S. Army Contracting Command (Dr. Daniel Cole) Cooperative Agreement (W911NF2420168)

**Authors contributions**:

Conceptualization: MK, SK

Methodology: KB, MK, LN, RK

Investigation: KB, MK, MP, VM, DC, SK, JP, RK, JZ

Visualization: KB, MK, SK

Funding acquisition: MK, SK, TŠ

Writing – original draft: MK, SK, KB

Writing – review & editing: all authors

**Competing interests**: The authors have no competing interests to declare.

**Data and materials availability**: The data underlying this study are openly available, see Ref. 57.

**List of Supplementary Materials**.

Methods and Modelling details

Figs. S1 to S16

Tables S1 to S4

References (44-56)

Movies S1-S8

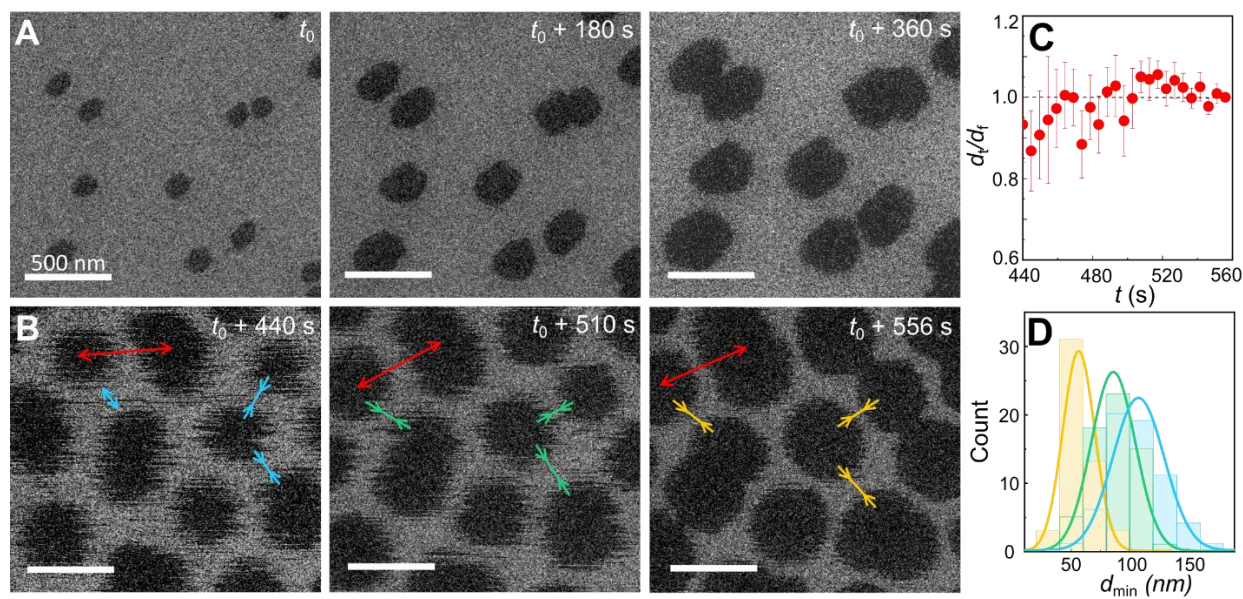

**Figure 1. Graphene growth on solid Au and self-assembly on molten Au.** Representative *in situ* scanning electron microscopy (SEM) images acquired from **A**) a solid Au sample at temperature $T = 1223$ K and **B**) after melting the same sample at $T = 1373$ K as a function of time $t$ during exposure to the ethylene ($C_2H_4$) gas at 30 Pa. We define $t_o$ as the total time elapsed ($\approx 3$ min) between the nucleation event and the first image shown in panel **A**. In the SEM images, graphene domains and Au substrate appear in a darker and lighter grey contrast, respectively. All the scale bars are 500 nm. **C**) Plot of the distances $d_t/d_f$ vs. $t$, where $d_t$ and $d_f$ are defined as the distances, respectively, between centers-of-mass of neighbouring domains at times $t$ (highlighted by red arrows in panel **B**) and at $t_f = t_o + 560$ s, the end time of this measurement sequence. **D**) Histograms of minimum distances $d_{min}$ between adjacent domains, indicated by arrows in panel **B** with cyan, green, and yellow colors in the plot corresponding to the times $t_o + 440$ s, $t_o + 510$ s, and $t_o + 556$ s, respectively. Solid colored curves are fits to the experimental data. Image panels **A** and **B** are extracted from movies, Movie S1 and S2, in the Supporting Material (SM), respectively.

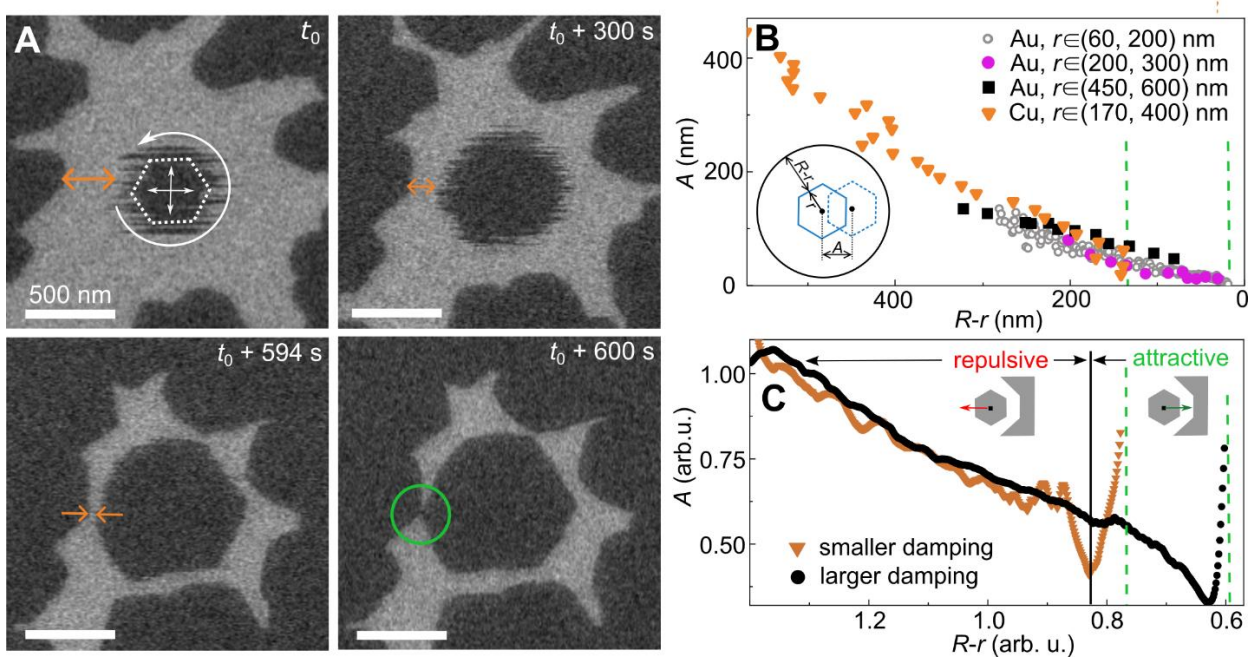

**Figure 2. Graphene domain oscillations on molten metal.** Panel **A** shows a series of *in situ* SEM images obtained during graphene growth on a liquid gold using 15 Pa $C_2H_4$ at $T = 1313$ K. In this experiment, the time between the substrate melting and the initial image in **A** is $t_o \approx 6$ min. The entire measurement sequence is available as Movie S3 in SM. The dotted hexagon in the first image in panel **A** highlights a wobbling domain with fuzzy contours. The arrows within the hexagon and curve arrow indicate translational and rotational motion of the domain, respectively. As the domain grows, it finally attaches to a neighbouring domain and the point of contact is highlighted using a green circle. **B, C**) Plots of experimental and simulated wobbling amplitudes $A$ vs. bare liquid spacing ($R$-$r$), indicated by the orange arrows in **A**. We define $r$ (= $\sqrt{S}$) as the orientation-averaged size of the domain of the area $S$ and $R$ as the distance between domain's center-of-mass to the surrounding boundary as illustrated by the schematic in **B**. The data extraction procedure is described in Methods and Fig. S4. The data plotted in **B** are obtained from graphene domains of sizes $r \in \langle 60, 200 \rangle$ nm (■), $\langle 200, 300 \rangle$ nm (●), and $\langle 450, 600 \rangle$ nm (■) growing on molten Au with 15 Pa $C_2H_4$ at $T = 1313$ K and of sizes $r \in \langle 170, 400 \rangle$ nm (▼) on molten Cu with $1.2 \times 10^{-2}$ Pa $C_2H_4$ at $T = 1373$ K. The $r$ values in parentheses correspond to the initial and final sizes of the growing domains during the measurements. In plot **C**, the black and orange curves are data extracted from simulations shown in Movies S4 and S5, respectively, carried out using larger and smaller damping coefficients. The solid black line marks the moment of the change from repulsive to attractive forces acting on the domain. The dashed green lines in **B** and **C** indicate the moment of attachment of the domains to other ones or to the enclosing boundary.

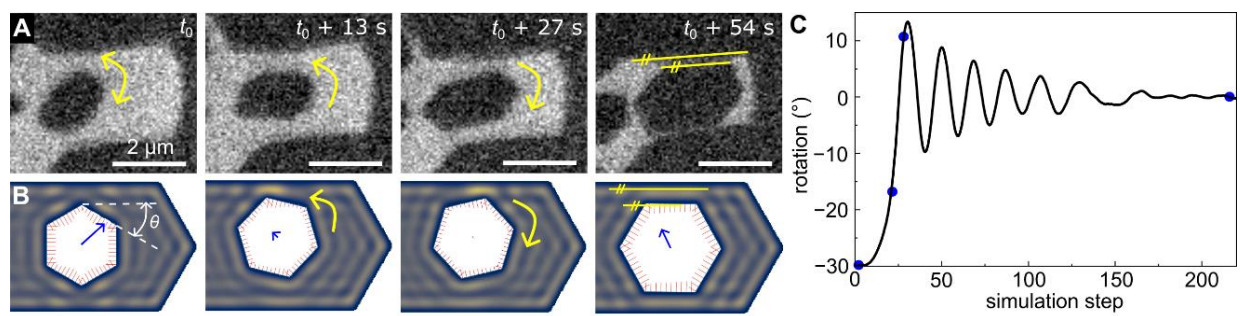

**Figure 3. Reorientation dynamics of graphene domains on molten metal.** Panel **A** shows typical *in situ* SEM images acquired from molten Cu at $T = 1373$ K during graphene growth using $1 \times 10^{-2}$ Pa $C_2H_4$. Here, the time since copper melting to the initial image in **A** is $t_o \approx 90$ s. (Full movie available as Movie S6.) SEM images reveal the growth and reorientation of an individual domain enclosed in a confined space. Yellow arrows show directions of the domain rotations. Within $t_o + 54$ s, the domain attains a stable position, which is with the longest edges parallel to the boundary, as indicated by the yellow parallel lines. Panel **B** shows a sequence of simulated images of a hexagonal domain floating on a liquid within nearly hexagonal compartment (please see Modelling for the details and a full movie of the simulation, Movie S4). Dark blue and yellow colors of the simulated liquid denote low and high amplitudes, respectively, of the waves emerging on the surface of the liquid. Lengths and directions of blue arrows represent magnitudes and orientations of the forces acting on the growing domain. Short red lines represent the forces at individual positions around the domain edges. **C**) Plot of domain orientation θ (defined in panel **B**) as a function of simulation time. Solid blue circles in the plot correspond to the images in panel **B**.

# Supplementary Materials for

**Casimir-like effect driven self-assembly of graphene on molten metals**

Kristýna Bukvišová[1,2], Radek Kalousek[1,3], Marek Patočka[3], Jakub Zlámal[1,3], Jakub Planer[1], Vojtěch Mahel[2,3], Daniel Citterberg[3], Libor Novák[2], Tomáš Šikola[1,3], Suneel Kodambaka[4], Miroslav Kolíbal[1,3]*

[1]CEITEC BUT, Brno University of Technology, Purkyňova 123, 612 00 Brno, Czech Republic

[2]Thermo Fisher Scientific, Vlastimila Pecha 12, 627 00 Brno, Czech Republic

[3]Institute of Physical Engineering, Brno University of Technology, Technická 2, 616 69 Brno, Czech Republic

[4]Department of Materials Science and Engineering, Virginia Polytechnic Institute and State University, Blacksburg, Virginia 24061, United States

Corresponding author: kolibal.m@fme.vutbr.cz

**The PDF file includes:**

    Methods and Modelling details
    Figs. S1 to S16
    Tables S1 to S4
    Description of supplementary movies S1-S8

**Other Supplementary Materials for this manuscript include the following:**

    Movies S1-S8

**Materials and methods**

Observation of graphene growth in a MicroReactor. All the graphene growth experiments on solid and molten gold are carried out using a MicroReactor (see fig. S1A) in a Thermo Scientific Helios 5 UC system equipped with a scanning electron microscope (SEM) and focused ion beam (FIB) milling. Gold samples (40 × 40 × 30 μm$^3$) were extracted from a 50 μm-diameter gold wire (Agar Scientific, 4N purity) *via* FIB milling in the Thermo Scientific Helios 5 Plasma FIB. In order to avoid contamination, gold was redeposited as a means to attach the chunk to the nanomanipulator. The gold samples were then positioned on the NanoEx MEMS heating chip and glued by ion beam-induced deposition. The chips were then placed in the MicroReactor and loaded into Helios 5 UC DualBeam equipped with a custom-built gas feeding system with individual inlets for dosing $C_2H_4$, $O_2$, and $H_2$ gases. Au pieces were first annealed in oxygen at temperatures *T* between 1073 and 1273 K for a variable amount of time (units of hours) to remove carbon deposits and other surface contaminants. Then, the Au was melted in ~10 Pa $H_2$ environment by heating above 1337 K. Any contaminants that emerged on the surface of the molten Au were then removed by gentle Ga ion beam milling in the FIB at room temperature. Finally, the samples were annealed sequentially in oxygen and hydrogen atmospheres until visually clean molten Au droplets are obtained. Graphene growth experiments are carried out by dosing ethylene ($C_2H_4$) gas at the desired temperature.

SEM imaging conditions. SEM images are acquired in the secondary electron mode using 10-20 kV accelerating voltages and electron beam currents between 50 pA and up to 1 nA to minimize charging. To test for the effects of electron-beam on the observed phenomena, graphene growth rates and graphene domain motion were compared during continuous and intermittent imaging. We did not observe any electron-beam-induced effects on the growth kinetics. The domain wobbling occurs independent of the beam parameters and while we cannot prove their existence with the beam off, we find that the domains remain in place without attachment to their surroundings whether the beam is on or off (see fig. S7 and S9).

Observation of graphene growth in UHV SEM. A modified Tescan ultrahigh vacuum (UHV, base pressure < 10$^{-6}$ Pa) SEM was used for all the experiments of graphene growth on solid and liquid Cu. Copper wire (EM-Tec, 3N purity) was wrapped around a V-shaped flattened platinum wire (Goodfellow, 4N purity) that was resistively heated by passing current in the UHV SEM. The temperature was monitored using a pyrometer (Micro-Epsilon thermometer CTLaser M3), calibrated by assigning Cu melting point to the temperature at which the melting was observed in the SEM. Melting was performed in 5×10$^{-3}$ Pa hydrogen atmosphere to mitigate oxidation of copper. Graphene is grown by slowly introducing ethylene after closing the hydrogen valve. Because of the low precursor pressures during the reaction, nucleation took place tens of minutes after the first introduction of ethylene, when the pressure was fairly stabilized. The molten copper evaporated rapidly, so the time available for observation of the liquid surface was limited. Furthermore, increasing the temperature above the melting point induced mixing of platinum substrate with the copper droplet (see discussion in SI). Imaging was performed at 5 kV with a beam current in the range 300 pA to 1 nA, with no observable effect of the electron beam.

Image processing. Growth of graphene domains and areal coverages are tracked using a Python script. First, the image signal is improved, by removing noise, using Block matching and 3D filtering (bm3d package). Background was removed using the Rolling ball background subtraction (part of scikit-image package). Finally, the image was thresholded using the Otsu's method and the segmentation was visually checked. For the size evolution of a specific domain, the segmentation was manually processed using the ImageJ software.

Analysis of the domain oscillation dynamics was carried out by a combination of automated script and manual processing. For each frame, we calculated the extrapolated radius of the domain based on the evolution of graphene growth rate surrounding the domain and its size after attachment. Further, we determined the envelope of the oscillatory motion, from which we derive its amplitude, and the size of the uncovered surface surrounding the domain. More details can be found below in the section titled: SEM image analysis: Data extraction procedure.

*In situ* high-temperature atomic force microscopy (HT-AFM). It is important to note that the *in situ* HT-AFM measurements are critical to assess the surface meniscus of the molten metal as *ex situ* characterization of graphene-metal interfaces at room-temperature can lead to misleading deductions (see fig. S16).

High-temperature topography measurements by AFM were performed using LiteScope AFM-in-SEM microscope with Akiyama self-sensing probe (resonant frequency ~45 kHz, spring constant ~5 N/m, tip radius <15 nm) in the frequency-modulated tapping regime in the high vacuum of Versa FIB-SEM. The sample (Au particle) was melted by the MEMS chip, graphene crystals were grown on top of it by CVD by introducing 65 Pa of ethylene into the SEM chamber for 5 minutes. After that, the ethylene was pumped, and the sample was kept under high vacuum in the molten state for the AFM measurement. Since the AFM tip, a solid, is likely at a lower temperature than the molten metal surface, prolonged tip-surface contact may lead to local solidification. Therefore, special precautions were taken to avoid this possibility. During the measurement, the probe was approached to the sample using a comparatively low setpoint of 2 Hz (commonly used values range from 5 to 20 Hz or even higher) in order to decrease the tip-sample force as much as possible. In some instances, the setpoint (i.e., the tip-sample force) was adjusted during scanning to reveal how the observed height of the step between liquid gold and graphene depends on the tip-sample force.

Density functional theory (DFT) calculation procedure. All DFT calculations were performed with the Vienna ab initio simulation package (VASP) (*44*) using the projector-augmented wave method (PAW) (*45*) for treating core electrons. Eleven valence electrons for gold and copper, four valence electrons for carbon, and one valence electron for hydrogen are described with a plane-wave basis set with an energy cut-off set to 450 eV. The Perdew-Burke-Ernzerhof functional was adopted for the exchange-correlation energy, supplemented by the Grimme pairwise dispersion correction D3(*46*). The structural and electronic calculations were performed in two steps: The gamma-point calculations were employed for initial optimization, and the process was terminated when the residual forces acting on ions decreased to values below 0.04 eVÅ$^{-1}$. Subsequent electronic structure calculations involved a Gamma-centered 2×2×1 Monkhorst-Pack grid (*47*) for Brillouin zone sampling, and the number of FFT grid points was increased by 30%.

Gold and copper surfaces were represented using 4-layered (111) slabs within a rectangular supercell with the matrix notation (11 0 | -6 12). The dimensions of the supercells were approximately 26.2×27.7Å and 30.1×31.9Å for the copper and gold surfaces, respectively. The periodically repeated sheets were separated by 15 Å thick vacuum layer and all calculations included dipole corrections to both the potential and energy. Bader charge analysis (*48*) was conducted to assess the charge transfer between a metal substrate and a graphene domain.

SEM image analysis: Data extraction procedure

The image sequencies were processed according to the description below, resulting in plot in fig. S4.

On liquid gold:
- **Domain radius** *r* was extrapolated back in time based on the growth rates of the domains that were static (e.g. pinned to a stationary edge). As soon as the domain became attached to the surrounding domains, we measured its size and used it for the retrospective analysis, based on the evolution of growth of the static graphene surrounding the oscillating domain. Secondly, the graphene was simply thresholded similarly to the above-mentioned approach for the coverage evolution. These two approaches produced nearly identical results, deviating only for the initial stages, when the domain is very small and oscillates vigorously.
- **Amplitude of motion**, *A*: For each frame, we measure the envelope of the smeared domain. We used two strategies, manual segmentation and automated image processing in a Python script. In the script, we used a Gaussian kernel 1x3 pixels that induced "smearing" in vertical direction and then the domain envelope was thresholded. There was a good agreement between the data extracted by both approaches, with the exception of initial stages of growth. There, the domain size was small and the smeared domain appeared discontinuous in the image, being unable to analyze by the automatic scripting. Therefore, the data shown is the one acquired manually.
- **Radius** *R* of the area in which the oscillating domain is enclosed: For each image in the sequence, we inscribed a largest circle in which the domain could oscillate without coming into direct contact with the surrounding domains. *R* is the radius of the inscribed circle.

On liquid copper:
- Thanks to the possibility to capture arbitrary scan windows, the frame-time was reduced and we did not analyze smeared domains like in the case of liquid-gold-related datasets. Instead, we are able to directly measure domain parameters in each frame (*r*,*R*). When large-area scan was preformed, the domain was smeared in an identical way as in the MicroReactor datasets.
- However, the time between two subsequent images is larger compard to those observed on molten Au, so we have relatively fewer data on the domain movement. The amplitude *A* was determined as the envelope of centre-of-mass evolution in time.

Domain displacement rates. The rates of domain motion are between $10^{-6}$ and $10^{-5}$ m/s, as deduced from the scanning rate of the electron beam, for both copper and gold surfaces, despite different beam scanning strategies. The rates are estimated to be in this interval by the following procedure. For the lowest dwell times possible (50-100 ns) and low magnifications, the 'free' domains appear sharp, meaning that the time it takes to scan over the particle is comparable or smaller than that required for the domain to move away from the scanned area. This gives an estimation for the upper boundary for the velocity interval. At longer dwell times, we observe displacement of the domain in between two consecutive scans. This displacement, together with the frame time, determines the lowest possible velocity of the domain.

**Modelling details**

Casimir-like surface undulations. We concentrate only on mechanical waves that appear on a free liquid surface due to thermal/mechanical vibrations of the whole system. We assume that these mechanical waves are the only cause of forces acting on the domains. Different waves can appear at different sides of the domain manifesting in both repulsive and attractive interaction between domains (see fig. S13). This interaction has an analogy in Casimir forces well-known in quantum field theory.

Let us consider the free liquid surface as a region where the mechanical waves can appear while the surface covered by the domain is completely solid (e.g. we treat the simplest case where elasticity of the graphene domains is neglected). The waves are described by a solution of the 2D Helmholtz equation (stationary wave equation) where the domain edges represent a Dirichlet boundary condition with zero displacement. The amplitudes of the waves with a specific wavelength (wavenumber) are given by a resonance curve (similar the theory of driven oscillations), see fig. S13. The boundary condition is a result of the wave reflection from the domain edge. Since every wave carries certain momentum (and of course energy), the wave reflection is accompanied by the change of momentum manifested in forces pushing perpendicularly against the domain edge. These forces have always the same direction regardless of the instant direction of the wave displacement near the edge, therefore, let us consider that the force is proportional to the square of the directional derivative of a wave at the edge point with respect to the direction perpendicular to the edge. The domain motion (both translational a rotational) is obtained by solving an equation of motion containing all forces along the entire edge.

A solid region where the domain(s) is/are considered to move is selected. This region represents the surrounding domains that are already immobile. The edges of this region are fixed and are defined by the same Dirichlet boundary condition (zero displacement) as well as the domain edge. The domain of a certain shape, size, and orientation is placed in the region. In the first step, the 2D Helmholtz equation

$$\Delta u_k + k^2 u_k = 0$$

for all wavenumbers $k$ is solved inside the region with the Dirichlet boundary conditions of zero displacement mentioned above. As expected, these solutions form a set of eigenfunctions belonging to specific eigenvalues of $k$. The amplitudes $A_k$ of every eigenfunction are chosen according to a unique and temporarily constant resonant curve defined by a specific position and width of its peak. The magnitude of the force $\vec{F}_i$ acting perpendicularly against the $i$-th point of the domain edge is proportional to the sum

$$\sum_k \left( A_k \, \partial_{\vec{n}_i} u_k \right)^2$$

where $\partial_{\vec{n}_i}$ is the directional derivative with respect to the direction $\vec{n}_i$ perpendicular to the edge. The translational equation of motion is given as follows:

$$m\vec{a} = -b\vec{v} + \sum_i \vec{F}_i,$$

where $m$ is the domain mass, $\vec{a}$ and $\vec{v}$ are the domain acceleration and velocity, respectively, and $b$ is the drag constant. The rotational equation of motion reads

$$I\vec{\varepsilon} = -\beta\vec{\omega} + \sum_i \vec{r_i} \times \vec{F_i},$$

where $I$ is the moment of inertia, $\vec{\varepsilon}$ and $\vec{\omega}$ are the respective domain angular acceleration and angular velocity, all with respect to the domain center of mass, $\beta$ is the angular drag constant, and $\vec{r_i}$ is the position of $i$-th point of the domain edge with respect to the domain center of mass. In the next step, these two equations of motion predict small changes of the position and orientation of the domain resulting in slightly different set of eigenvalues $k$ and eigenfunctions $u_k$ when solving the 2D Helmholtz equation for the new scenario. The numerical simulation of the domain motion consists in repeating these two steps in which a gradual growth of the domain can be taken into account as well.

Further, we describe the meaning of used quantities and functions. Displacements $u$ refer to surface waves that origin in thermal fluctuations (where amplitudes of several Angstroms have been reported in experimental works (*35*, *49*). Additionally, the displacements may be related to the effects of adsorption/desorption kinetics within a smooth transition region between the liquid and vapor phase, which can have such short wavelengths (~ 100 nm). The damping constants $b$ and $\beta$ are principally related to the dynamic viscosity of the liquid metal, resulting in losses in the mechanical energy of the domain. The resonant curve (*50*, *51*) results from the following consideration: the liquid behaves as a resonant system exhibiting a certain resonant frequency. This system is driven by intrinsic thermal and mechanical vibrations at various frequencies causing high amplitudes of oscillations at frequencies close to the resonant one, while very low amplitudes at different ones.

Estimation of the charge transferred to the graphene. We have used Density Functional Theory to quantitatively assess the charge density redistribution between a graphene flake and the metal substrate. To elucidate whether the graphene flakes are H-terminated (fig. S11A-C) or metal-terminated (fig. S11D-F), we compare the calculated binding energy of the hydrogen atoms with the hydrogen chemical potential at experimental conditions ($\mu_{H2}$=-2.08 eV for 1070 °C and $10^{-5}$ mbar $H_2$). The binding energy of the hydrogen atoms terminating a graphene flake is calculated to be -1.89eV, i.e., close to the hydrogen chemical potential. Therefore, we conclude that DFT calculations do not exclusively prefer any of the terminations considered. The binding energy of hydrogen in the graphene flake on the copper substrate is reduced to -1.57 eV, showing that DFT prefers the Cu-terminated graphene flakes. Since the metal-terminated flakes are calculated to be less charged than the flakes terminated with hydrogen, as discussed later, we will build a phenomenological model for the estimation of the charge transfer between the Au(111) and Cu(111) substrates and a hydrogenated flake. Here we assume that graphene domains interact via electrostatic dipole-dipole interaction. Two effects contribute to the size of the induced dipole moment at the interface between a graphene domain and a substrate. The first, (fractional) charge transfer, is caused by different work functions of the metal and the graphene, while the second one, known as the push-back effect, is induced by the electrostatic repulsion, where the graphene layer pushes a metallic charge density spillover back to the substrate.

To estimate the charge density redistribution upon adsorption in the graphene layer, we employ the modified Bader charge analysis: The Bader volumes are evaluated from the all-electron charge density of the system on a denser FFT grid. Then, we integrate the charge density difference $\Delta\rho(x,y,z)$ between the interacting and separated systems over the Bader volumes. Specifically, $\Delta\rho(x,y,z)$ is calculated as

$$\Delta\rho(x,y,z) = \rho_{substrate+flake} - \rho_{substrate} - \rho_{flake}.$$

To assess the validity of this approach, we compare the resulting total charge transferred from a substrate to a flake with the maximum of the cumulative charge transfer function $Q(z)$, which contains contributions from both the fractional charge transfer and from the push-back effect. This is defined as

$$Q(z) = \iint \int_{-\infty}^{z} \Delta\rho(x,y,z)\,dz.$$

Calculated charge density differences and cumulative charge transfer functions for the graphene flakes on the Au(111) substrate shown in fig. S12 reveal the electron accumulation close to the substrate layer and electron depletion at the graphene flake. Qualitatively same behaviour is also observed for the Cu(111) substrate.

Table S1 summarizes the total charge transferred between the substrate and hydrogen-terminated graphene flakes. As both approaches produce comparable results, with relative differences below 2.4% for Au(111) and 11.2% for Cu(111), we conclude that the modified Bader charge analysis is a reliable method for evaluating the distribution of transferred charge on the flake.

Next, we define a surface charge density $\sigma(R)$ of the redistributed charge inside the flake upon adsorption on a substrate using the calculated Bader charges. Independent of the size, the graphene flake can be divided into inner and outer regions where the transferred charge scales linearly with the number of atoms therein, only with different slopes, as shown in figure S13B,C for the Au(111) and Cu(111) substrates, respectively. Therefore, we assume that the inner region has the surface charge density $\sigma_i$ and the outer region $\sigma_o$. Also, the width of the outer region is fixed to $d$ for all flakes independent of their sizes. Thus, $\sigma(R)$ can be defined for all calculated flakes as:

$$\sigma(R) = \sigma_i \cdot \theta[(R_T - d) - R] + \sigma_o \cdot \theta[R - (R_T - d)],$$

where $R_T$ is the radius of the flake and $\theta(x)$ denotes the Heaviside function. Integration of $\sigma(R)$ over the area, shown as dashed lines in Figures S13D,E, yields the total charge transferred to the pink circle depicted in Figure S13A. This can also be expressed as a function of the number of atoms inside the circle, illustrated by the dashed lines in Figures S13B,C under the assumption that the surface density of carbon atoms agrees with graphene, i.e. ~2.6 Å$^2$ per carbon atom.

Overall, three parameters were fit to reproduce the charge transfer obtained from the modified Bader analysis (points in Figures S13B-E): the surface charge density of a transferred charge in the inner and outer regions of the flake $\sigma_i$, $\sigma_o$, and the width of the outer region $d$. The optimal parameters for all the flakes and for the graphene sheet adsorbed on both Au(111) and Cu(111) substrates are summarized in table S2. There is only one set of parameters needed for all calculated flakes on the Au(111) substrate, with $\sigma_i$ deviating by 13% from graphene. The flakes adsorbed on the Cu(111) substrate converge slower with $\sigma_i$ to graphene. Therefore, two sets of parameters are necessary: one for the smaller 3×3 and 4×4 flakes and one for the 5×5 flake, which differs in $\sigma_i$ by 4% relative to graphene.

Finally, we evaluate the position of the mirror charge by fitting the resulting dipole moment to the value obtained from the cumulative charge transfer function. The optimal separation distance is calculated to be 1.85Å for Au(111) and 1.72Å for Cu(111), with relative errors below 6%. These values are in agreement with the calculated plane-averaged charge density differences shown in figure S12. Furthermore, the dipole moment calculated from the cumulative charge transfer function is in perfect agreement with the compensating dipole moment of the SCF run, shown in table S3. Therefore, we conclude that there are no intrinsic dipoles present either in the slab or in the graphene flake.

This model is only valid for the hydrogen-terminated flakes, which do not contain intrinsic dipole moments. However, the metal-terminated flakes do not fulfil this assumption because of the significant bending observed in all structures (figures S11D-F). Instead of building a more complex model for the metal-terminated flakes, we show in table S4 that the resulting dipole moments from the SCF runs are always smaller than for the hydrogen-terminated variations. This behaviour can be explained by an increased charge transfer from the substrate to the layer arising from the compensation of the missing hydrogen bonds, which tends to counterbalance the push-back effect. Therefore, the metal-terminated flakes are expected to interact weaker than hydrogen-terminated flakes.

In summary, charge transfer and formation of the induced dipole for the hydrogen-terminated graphene flakes can be described by a constant surface charge density $\sigma_i$ over the whole area of the flake except for the 2.25Å thick edge region with the charge density 1.7 - 2.2× higher. The values of $\sigma_i$ presented in table S2 are for the largest calculated graphene flakes in a good agreement (relative errors below 13%) with the calculated surface charge density for the graphene layer. The charge transfer to the graphene flake is accompanied by a mirror charge, separated by 1.85Å and 1.72Å on Au(111) and Cu(111) substrates, respectively. These values were chosen such that they reproduce the total dipole moment in the system obtained from the SCF run.

Electrostatic (repulsive) interaction. As a result of charge redistribution, interface electric dipoles develop on each graphene domain and the domain-domain interaction can be treated as dipole-dipole interaction. This interaction is repulsive (dipoles pointing in the same direction) and has been previously considered to be of key importance in the alignment of 2D materials on LMCATs (*18*). We have conducted a calculation of electrostatic energy for a simplified geometry of two identical circular graphene domains floating on a flat metal surface. The electrostatic energy is calculated by integrating over four objects two graphene domains (indices 1 and 2) and two mirror charges inside the substrate (indices 3 and 4):

$$E_{el} = 2 \int_1 \int_2 \frac{1}{4\pi\varepsilon_0} \frac{dQ_1 dQ_2}{|\mathbf{r}_1 - \mathbf{r}_2|} + 2 \int_1 \int_4 \frac{1}{4\pi\varepsilon_0} \frac{dQ_1 dQ_4}{|\mathbf{r}_1 - \mathbf{r}_4|}.$$

The charge distributions which we have considered, are either constant over the graphene domain, or nonhomogeneous with highest values towards the edges (to cover both possibilities, due to uncertainty in DFT prediction for large domains). Another variable is the scaling of total charge – we have considered linear and quadratic scaling with radius. The linear scaling is reasonable for the charge concentrated at the edges, while the quadratic would be appropriate for charge uniformly distributed on the graphene surface.

Having calculated electrostatic energies, we have compared them first to thermal fluctuations $E_{thermal} = k_B T$, which may enforce attachment of otherwise repulsive domains. The distance of the domains was sought for at which the electrostatic and thermal energy match. This is then compared with the minimum distances of the domains measured from the experimental data (e.g. (Fig. 2B, last image in the sequence).

Recalling the experiments, the domains oscillate, keep certain distances from each other but do not attach to each other until a certain point. From the experimental in-situ observations we have deduced amplitudes of the wobbling motion and minimum distances between individual domains for a wide set of geometries. Interestingly, the amplitudes and minimum distances seem to be independent on the domain sizes (fig. S10) and instead strongly scale with the width of the uncovered area between the domains (Fig. 2B). This behaviour is in striking contrast with the results received from electrostatic potential calculation results. If the actual size of the domains does not matter, as is suggested by results shown in Fig. 2B and fig. S10, and the only defining parameter is the separation of the domains, then the charge on the domain (and, hence, the electrical dipole) has to be constant, independent on the domain size, in case of dominance of electrostatic forces. This is clearly not the case.

Further considerations accounting for other interactions. Apart from Casimir-like surface undulations, we have been soliciting for electrostatic dipole interaction (repulsive), van der Waals forces (attractive), and capillary forces.

van der Waals forces. In order to model the interaction, we attempted to add an attractive van der Waals component. To estimate the magnitude of van der Waals forces which could act as the attractive counterpart of repulsive electrostatic force mentioned above, we use the formula describing the forces acting between two identical (circular) graphene domains of radii *R*, derived in (*52*):

$$F_{vdW} = \frac{-15 A_H h^2 \sqrt{R}}{512 D^{7/2}},$$

where $A_H \sim 1 \times 10^{-19}$ J is the Hamaker constant, $h = 0.34$ nm is the thickness of the graphene layer (interlayer distance in graphite) and *D* is the separation distance between the two domains. If we consider an example geometry for which we have observed the oscillations, that is discs of $R = 200$ nm, separated by distance D = 50 nm, we obtain an estimation for the magnitude of vdW forces $F_{vdW} \cong -9 \cdot 10^{-18}$ N, a significantly weaker force compared to electrostatic one. Fig. S12 shows the dependence of the calculated vdW potential on the distance between the domains, further illustrating that van der Waals forces are too weak to cause attraction of the domains at the distances observed in experiment.

Capillary forces. Previously, capillary forces have been speculated to induce a nanoscale 'Cheerios effect' (*18, 53*), which is observed for floating macroscopic objects. However, specifically for monolayer graphene, there is no liquid meniscus around the domain, as evidenced by our in-situ measurements (Fig. 1, 2 and fig. S2,3).

**Table S1.** Calculated number of electrons transferred from the hydrogenated flakes to the Au(111) and Cu(111) substrates using the modified Bader charge analysis and the cumulative charge transfer function. The relative differences of the transferred charge are below 2.4% for the Au(111) substrate and 11.2% for the Cu(111) substrate.

|  | Au(111) | | | Cu(111) | | |
|---|---|---|---|---|---|---|
|  | 3×3 | 4×4 | 5×5 | 3×3 | 4×4 | 5×5 |
| Modified Bader | 0.67 | 1.04 | 1.47 | 0.80 | 1.33 | 1.63 |
| Max(Q(z)) | 0.66 | 1.03 | 1.48 | 0.72 | 1.21 | 1.58 |
| Relative differences (%) | 2.4 | 1.6 | -1.0 | 11.2 | 10.1 | 3.0 |

**Table S2.** Fitted parameters of our model to describe the charge transferred to the graphene flakes with varied sizes and to the graphene sheet adsorbed on Au(111) and Cu(111) substrates: Charge density of the inner and outer regions ($\sigma_i$, $\sigma_o$) and the width of the outer region $d$.

|  | Au(111) | | | | Cu(111) | | | |
|---|---|---|---|---|---|---|---|---|
|  | 3×3 | 4×4 | 5×5 | graphene | 3×3 | 4×4 | 5×5 | graphene |
| $\sigma_i$ (e$^-$Å$^{-2}$) | $-2.38 \cdot 10^{-3}$ | | | $-2.1 \cdot 10^{-3}$ | $-3.5 \cdot 10^{-3}$ | $-2.9 \cdot 10^{-3}$ | $-2.8 \cdot 10^{-3}$ | |
| $\sigma_o$ (e$^-$Å$^{-2}$) | $-5.24 \cdot 10^{-3}$ | | | - | $-6.0 \cdot 10^{-3}$ | $-4.9 \cdot 10^{-3}$ | | - |
| $d$ (Å) | 2.25 | | | - | 2.25 | | | - |

**Table S3.** Calculated dipole moments induced by the interaction of the graphene flakes with the Au(111) and Cu(111) substrates. Values obtained from the cumulative charge transfer function are in perfect agreement with the compensating dipole moment of the SCF run.

|  | Au(111) | | | Cu(111) | | |
|---|---|---|---|---|---|---|
| Flake size | 3×3 | 4×4 | 5×5 | 3×3 | 4×4 | 5×5 |
| Cum. charge transfer function (e$^-$Å) | -1.21 | -1.83 | -2.45 | -1.37 | -2.27 | -2.75 |
| SCF dipole corrections (e$^-$Å) | -1.22 | -1.83 | -2.48 | -1.39 | -2.29 | -2.75 |
| Model (e$^-$Å) | -1.15 | -1.82 | -2.60 | -1.37 | -2.27 | -2.75 |

**Table S4.** Comparison of dipole moments between the hydrogen-terminated and metal-terminated graphene flakes with varying sizes upon adsorption on Au(111) and Cu(111) substrates. The resulting dipole moments for metal-terminated flakes are smaller compared to the hydrogen-terminated flakes.

|  | Au(111) | | | Cu(111) | | |
|---|---|---|---|---|---|---|
|  | 3×3 | 4×4 | 5×5 | 3×3 | 4×4 | 5×5 |
| H-terminated (e$^-$Å) | -1.22 | -1.83 | -2.48 | -1.39 | -2.29 | -2.75 |
| metal-terminated (e$^-$Å) | -0.16 | -0.46 | -0.89 | -0.25 | -0.19 | -0.79 |

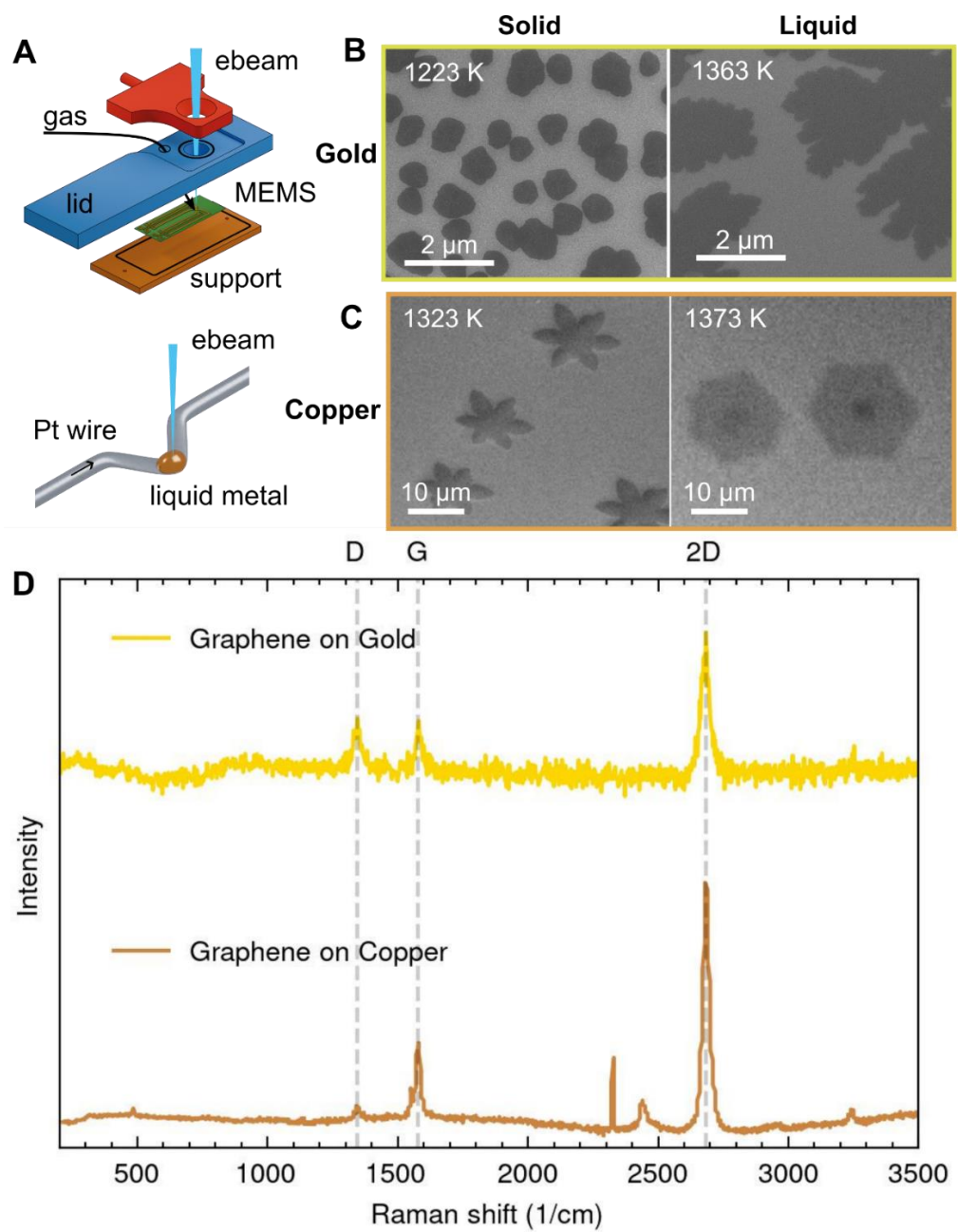

**Fig. S1. Domain shapes and Raman spectra of graphene**. (**A**) Schematics of the MicroReactor design (top) and resistively heated Pt wire support for the growth of graphene (bottom), respectively, on Au at high fluxes and on liquid Cu droplets at low fluxes in UHV SEM. (**B,C**) *In situ* SEM images of graphene domains acquired from Au (**B**) and Cu (**C**) during exposure to ethylene ($C_2H_4$) at 30 and $1.2 \times 10^{-2}$ Pa for gold and copper respectively on solid and liquid surfaces at temperatures *T* indicated in the panel. (**D**) Raman spectra of graphene grown on Au and Cu, obtained after rapid (50 K/s) cooling of the as-deposited samples to room-temperature. Higher intensity D peak in the Raman spectra from graphene on Au is likely a result of strain in the graphene layer on the droplet caused by rapid solidification. Laser power used was 0.2 mW and grating of 2400 g/mm.

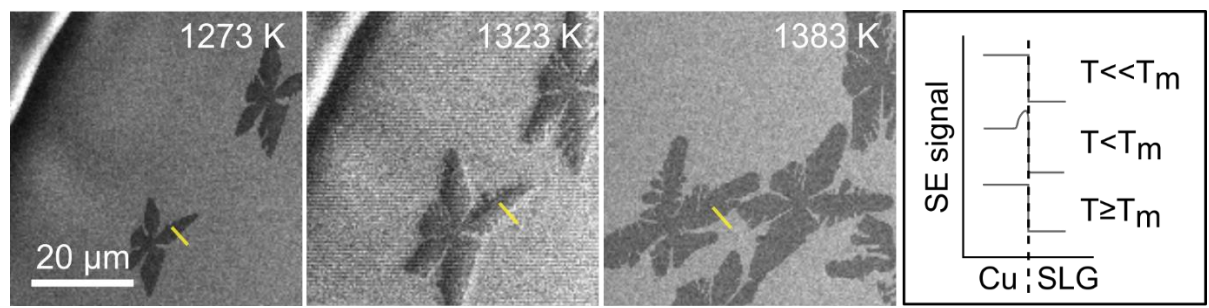

**Figure S2. Representative secondary electron (SE) images of graphene domains grown on solid and liquid Cu at different temperatures**. In this experiment, graphene is grown at $T = 1273$ K with $p_{C_2H_4} = 1 \times 10^{-2}$ Pa for approx. 8 min before the initial image in the sequence. SEM images do not show any 'rim' around the graphene domain edges on solid Cu substrate (left, 1273 K). At $T = 1323$ K, slightly below the melting point $T_{m, Cu}$ of Cu, a bright rim appears around the domains. The rim disappears when the substrate is molten at $T = 1383$ K. Schematic on the right illustrates the SE signal, in which SLG refers to single layer graphene. Secondary electrons are very sensitive to the curvature of the emitting surface (observation of single atomic steps is possible (*54*)), hence, any nanometer-scale irregularity of the surface is enhanced in SE image. The rim formation can be explained by pre-melting of the substrate localized to below the graphene domains only and has been observed before (*55*). Alternatively, the substrate could deform due to different thermal expansion coefficients of solid (surrounding the graphene) and liquid (below the graphene).

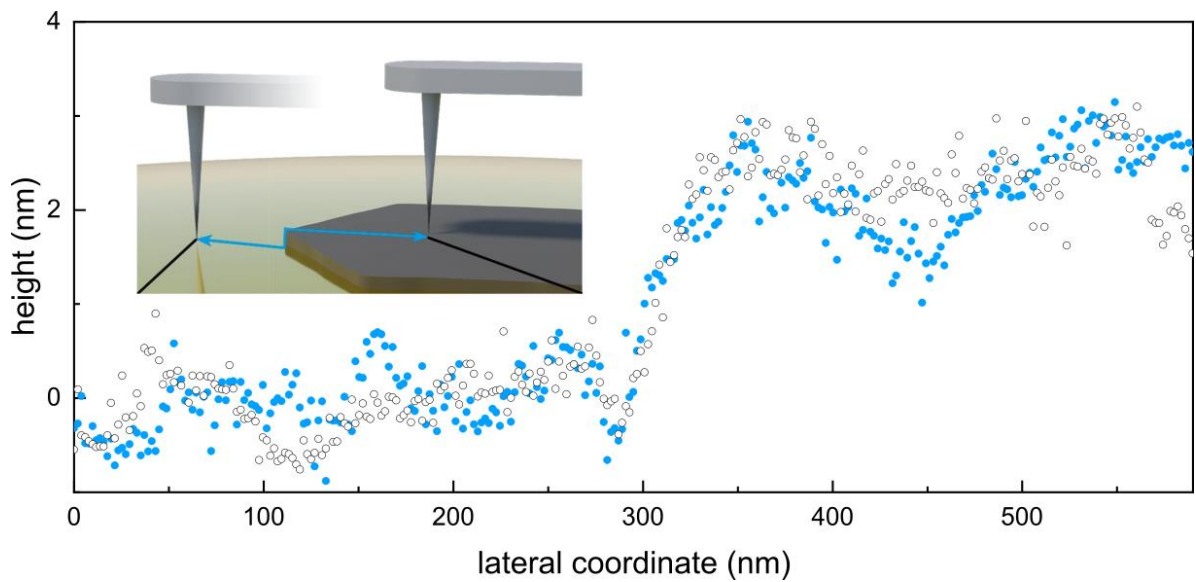

**Figure S3.** *In situ* **high-temperature AFM measured topography of a floating graphene domain on molten Au**, measured at 1343 K in high vacuum ($10^{-4}$ Pa) and a tapping mode. We chose a graphene domain that was pinned to other domains and, hence, stable during the measurement. The AFM image and the surface profiles (trace (○) and retrace (●)) do not show a clear view of the existence of a possible meniscus on the liquid surface. The inset shows a schematic of the measurement geometry.

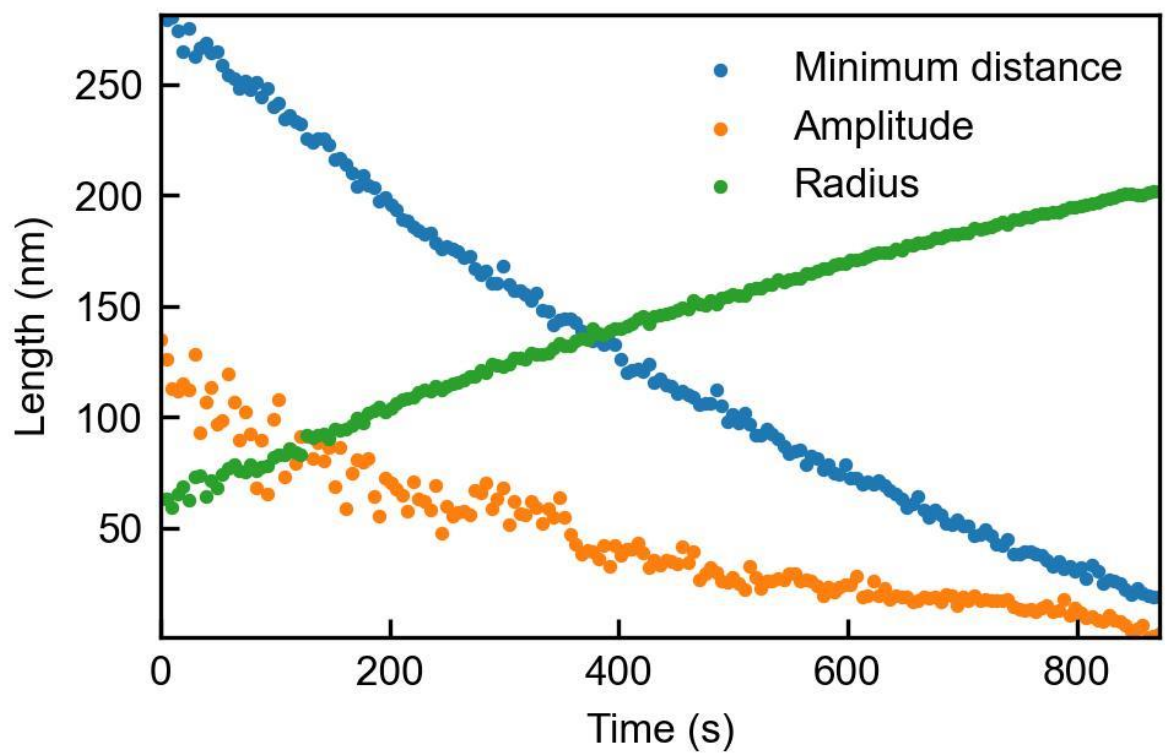

**Figure S4. Quantification of the image sequencies of oscillating domains**. Several quantities can be drawn from the image sequences, as described above. Analysis of the oscillating domains varied between the Cu and Au datasets because of varying frame rates, inherent to different microscopes that were used to capture the growth (see above).

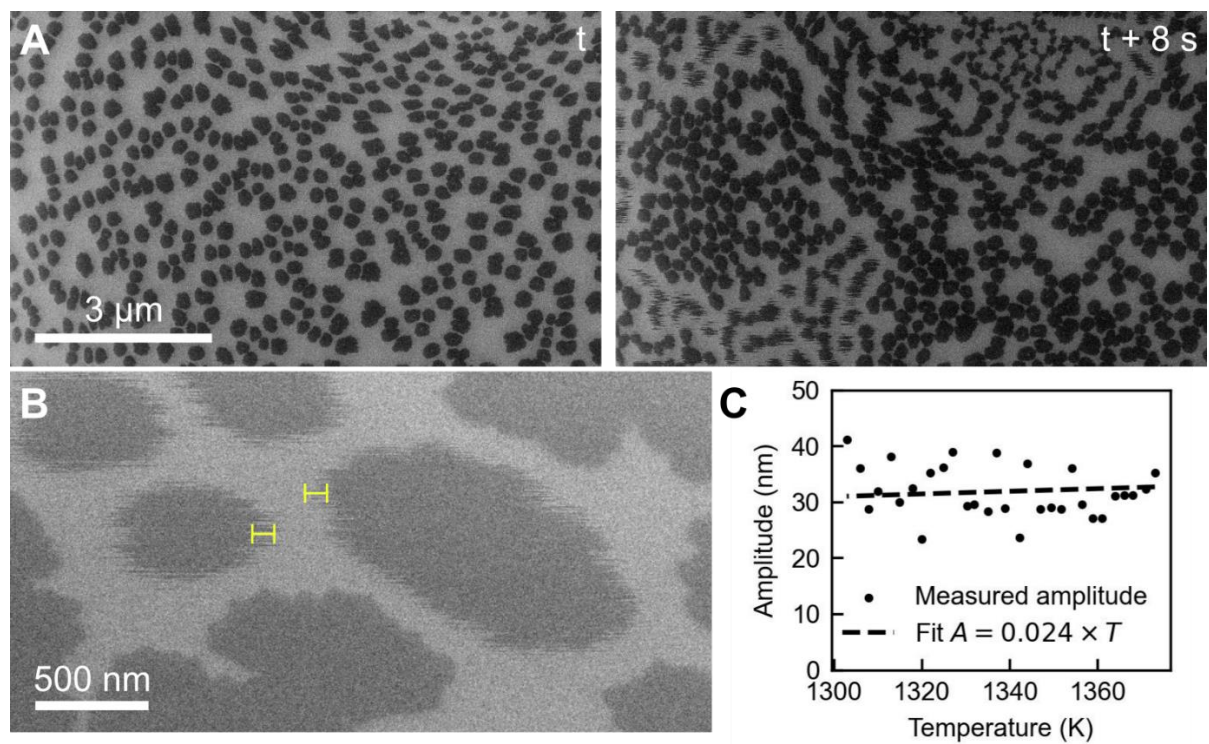

**Figure S5. Effect of temperature and domain size on domain dynamics:** *In situ* SEM images acquired from Au sample during the CVD of graphene using 35 Pa of ethylene. In **A**, the graphene domains were nucleated on solid Au at 1223 K (left) and then Au rapidly melted and undercooled back to 1223 K (right, after 8 s). Many domains immediately cluster together, and these agglomerates are pinned to Au droplet edges. Those domains left floating are oscillating on a liquid surface. A careful inspection of the figure reveals relatively large floating domains that oscillate as well; this is clearly demonstrated in **B**. **C)** Amplitude of the oscillations exhibits a statistically insignificant dependence on temperature within the accessible temperature range. Brownian motion (Einstein's equation) scales inversely with the size and depends linearly on kT (*56*). None of that is seen here. These data also support Fig. 2B, as they clearly show that the oscillations are not correlated to any other system variable (e.g. to the domain size).

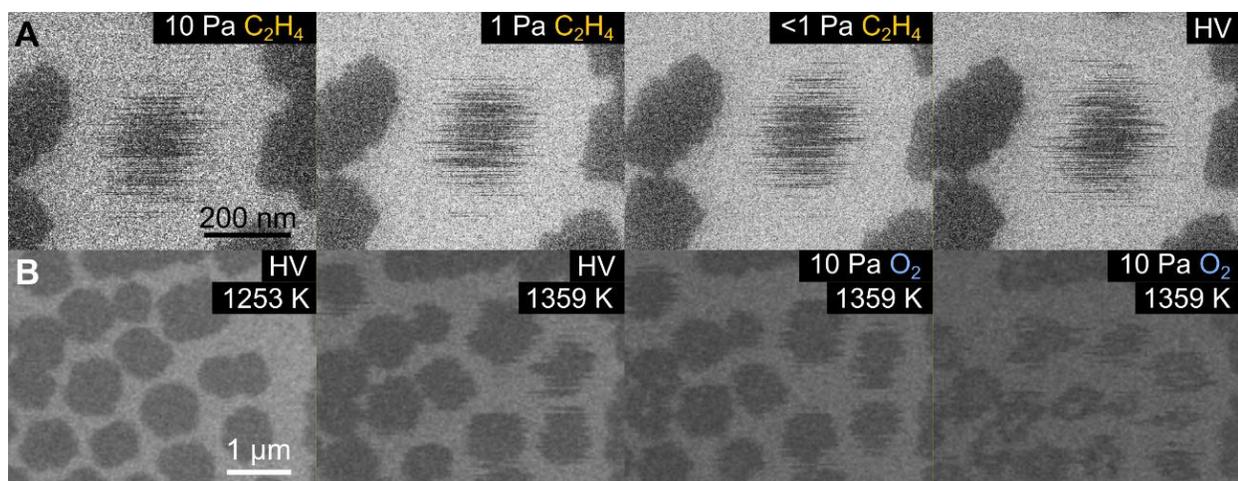

**Figure S6. Effect of the deposition flux on domain dynamics:** *In situ* SEM images from a Au sample during the CVD of graphene at different partial pressures of ethylene. The domain oscillations are independent of pressure. In **A**, the sample temperature is 1353 K, the pressure is noted in the images. HV = high vacuum ($10^{-3}$ Pa). In **B**, the first image is obtained for solid Au at 1253 K, while the subsequent frames are acquired at 1359 K. The domain oscillations are present also in $O_2$ gas, which results in etching of the domains. That is, the domains keep oscillating during graphene growth, in vacuum, and during etching as well.

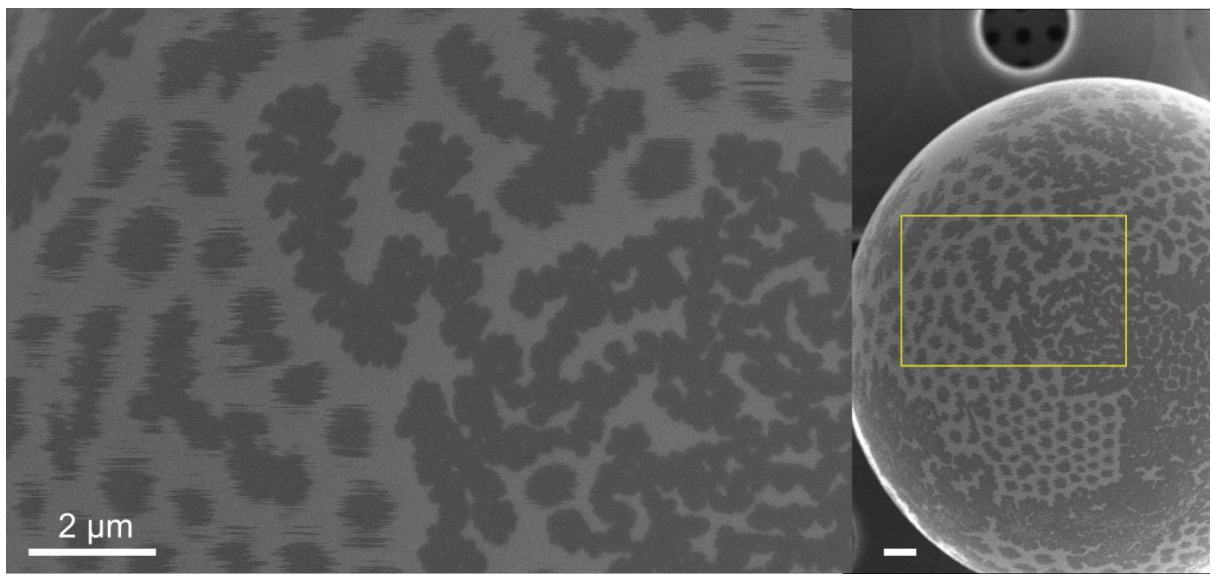

**Figure S7. Effect of electron beam on domain dynamics:** *In situ* SEM images acquired from Au sample during the CVD of graphene at an ethylene pressure of 15 Pa and 1311 K. SEM imaging conditions: 10 kV, 200 pA, dwell time 5 μs, electron beam dose rates: (left) 0.17 pA/nm$^2$ and (right) 0.01 pA/nm$^2$. We were imaging the yellow framed area for about 300 s (5 minutes) and observed the floating domain oscillations. The right image is a 4× demagnified image under the same beam conditions (hence, 16× lower electron dose), we see previously unexposed areas where floating domains exhibit identical oscillation behavior as in the previous scanning window. The oscillations, however, occur under different beam conditions and the domains behave similarly even if the beam is scanning a different area. Moreover, imaging under different beam conditions (including also a different microscope) does not affect the data plotted in Fig. 2B, where all the curves fall onto each other despite being taken under different beam conditions.

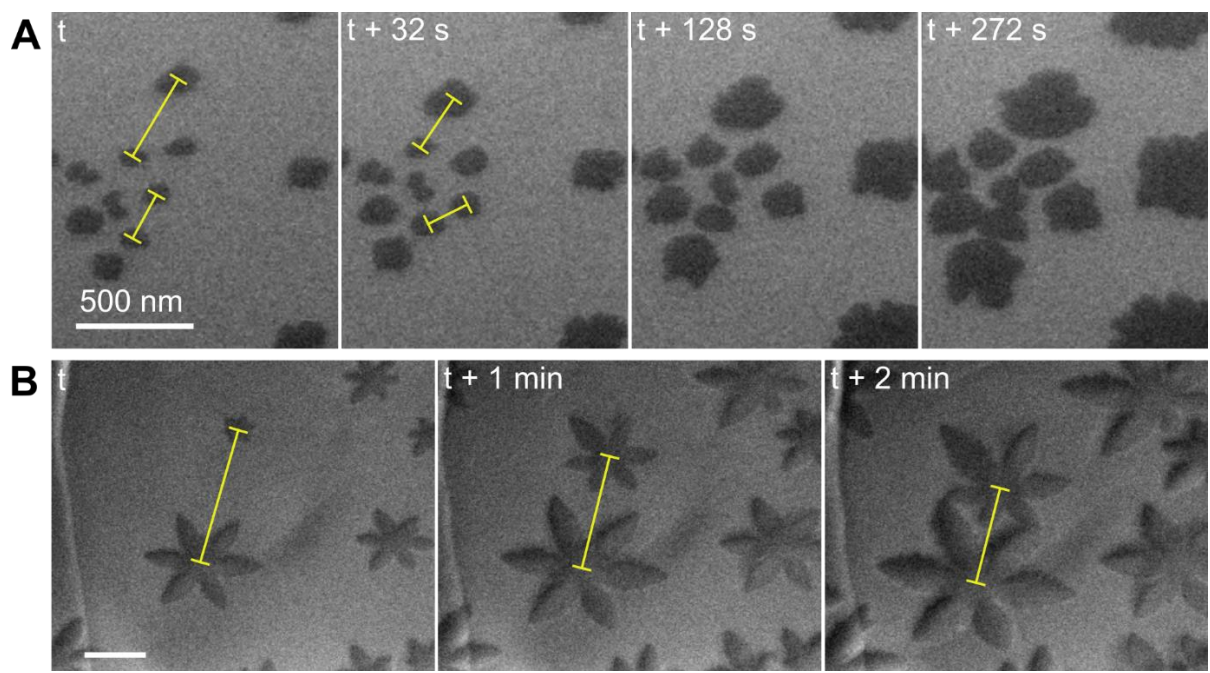

**Figure S8. Effect of the state of the substrate on domain dynamics and wobbling:** *In situ* SEM images from premolten Au and Cu samples during the CVD of graphene. The motion of the graphene domains is observed also on the premolten substrate (slightly before reaching the bulk melting point), although it is suppressed to some extent. However, the domains do not exhibit wobbling. The yellow line marks the shortening distance between the domains. A) Premolten Au, $T = 1323$ K, ethylene pressure 5 Pa, B) Premolten Cu, $T = 1343$ K, ethylene pressure $2 \times 10^{-2}$ Pa.
39

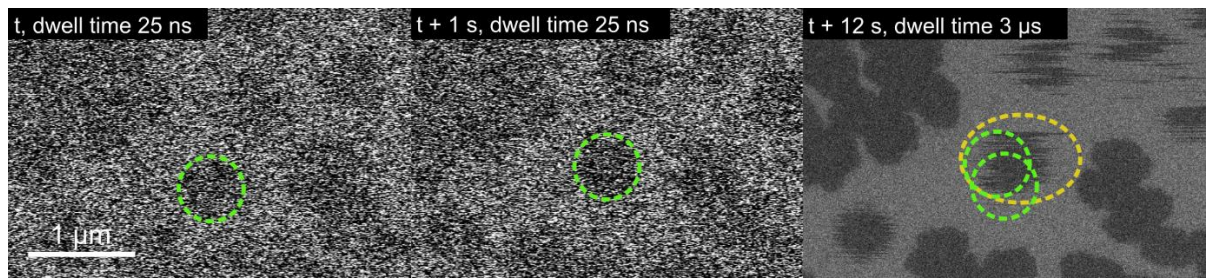

**Figure S9. Effect of dwell time on domain dynamics:** *In situ* SEM images of an undercooled Au sample during the CVD of graphene acquired using different dwell times. Imaging of unattached graphene islands on undercooled gold at different dwell times reveals that the oscillations seen in SEM images are related to domain movements, not to changes in shape (apart from increasing in size) – centre of mass (deduced from the short dwell time image, green circles) falls into the envelope (yellow ellipse) of oscillations measured on the larger dwell time image. Beam energy 10 kV, current 0.8 nA, temperature 1223 K, ethylene pressure 10 Pa.

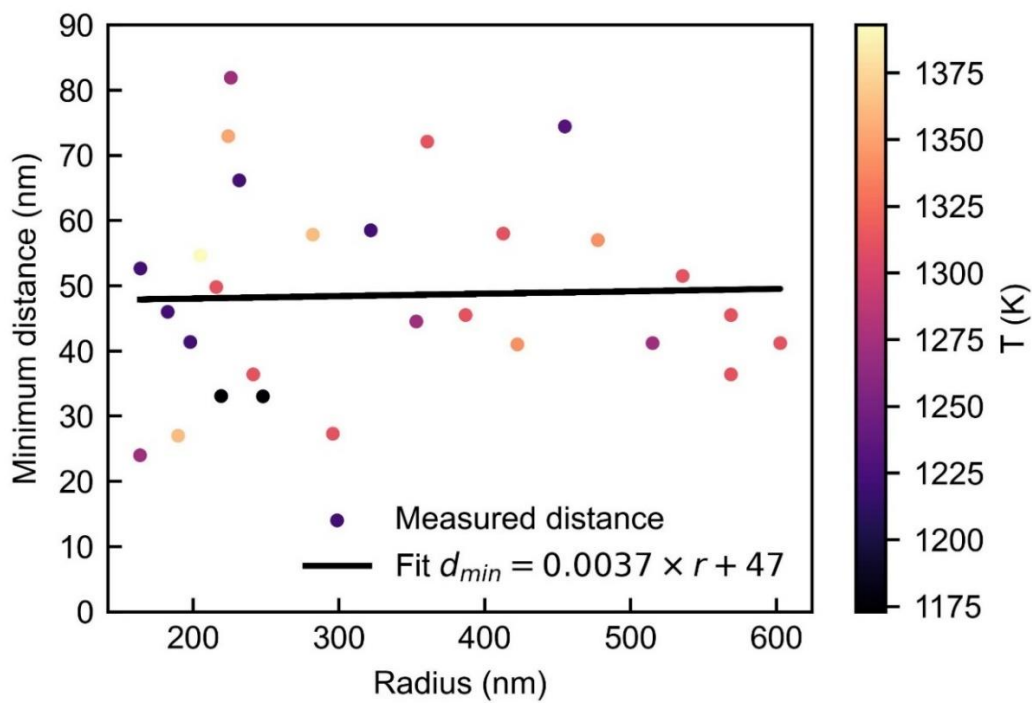

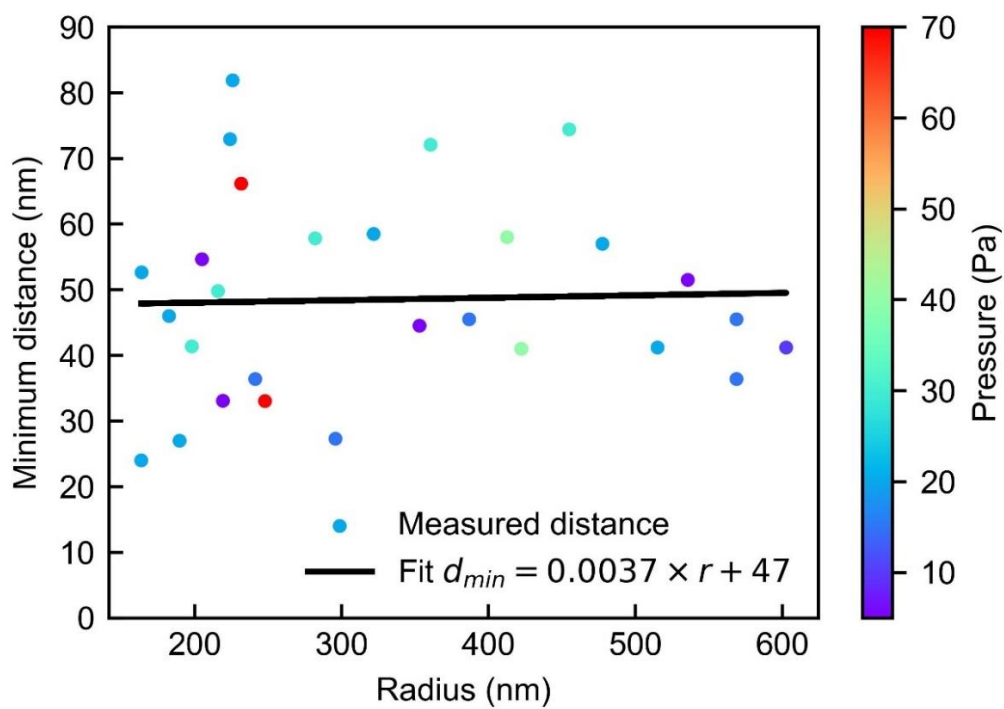

**Figure S10. Analysis of the critical distance as measured on liquid gold for a variety of experimental conditions (1223 – 1373 K, 5-70 Pa ethylene) and domain sizes**. Only a very weak dependence of the minimum distance on the sizes of domains is seen. Each datapoint represents a different experiment, the color code stands for the sample temperature (top) or $C_2H_4$ pressure (bottom). No correlation is found between the minimum distance and experimental conditions.

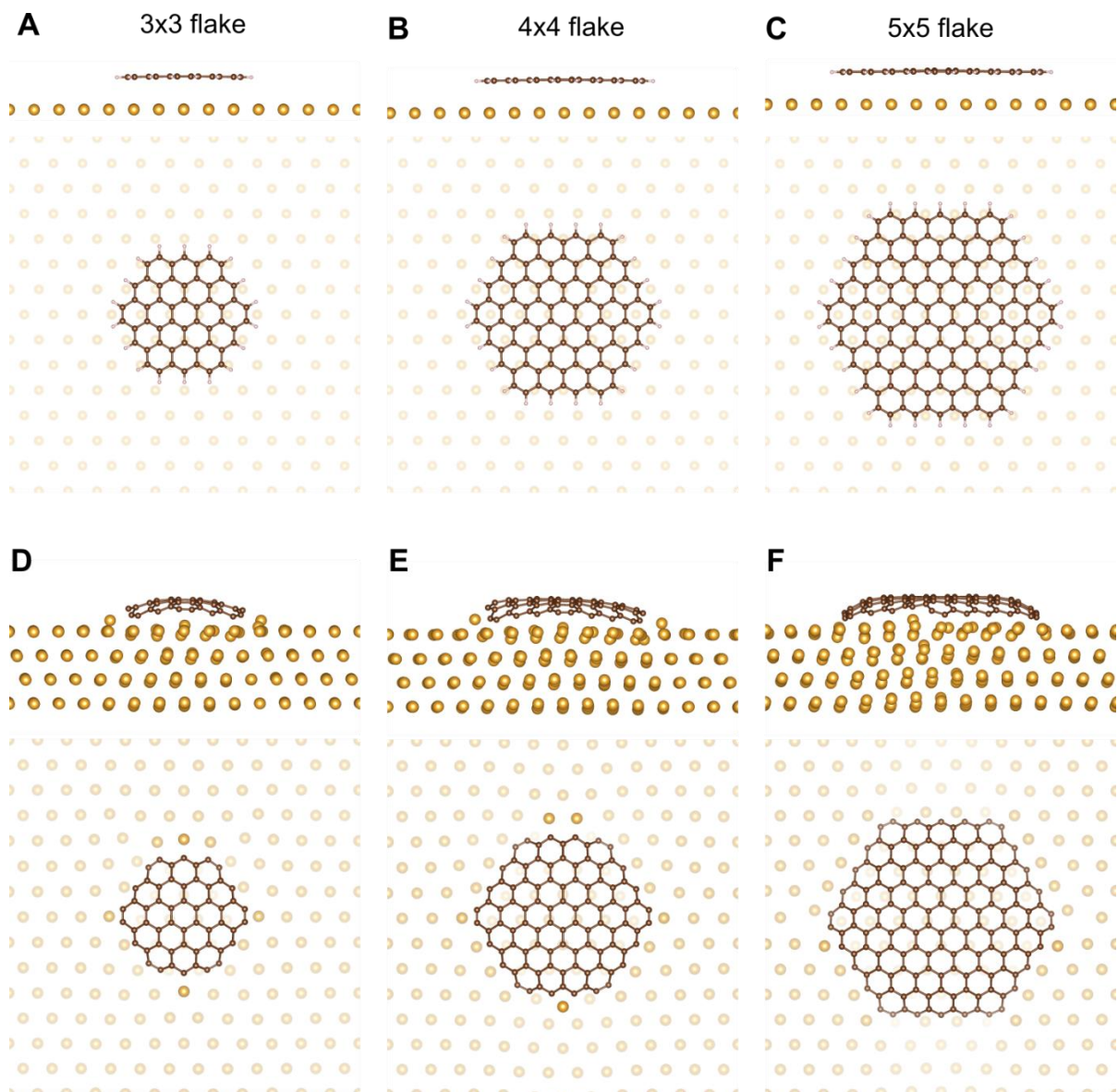

**Fig. S11**: **Top and side views on the relaxed geometries of the hydrogen-terminated (panels A-C) and Au-terminated (panels D-F) graphene flakes**. The calculated structures are hexagonally shaped with the side composed of 3 aromatic rings marked as 3×3 (panels **A** and **D**) up to 5 aromatic rings marked as 5×5 (panels **C** and **F**). Au-terminated flakes are bent toward the substrate, causing significant buckling in the first layer of gold. Qualitatively similar structures were also obtained for the Cu(111) substrate.

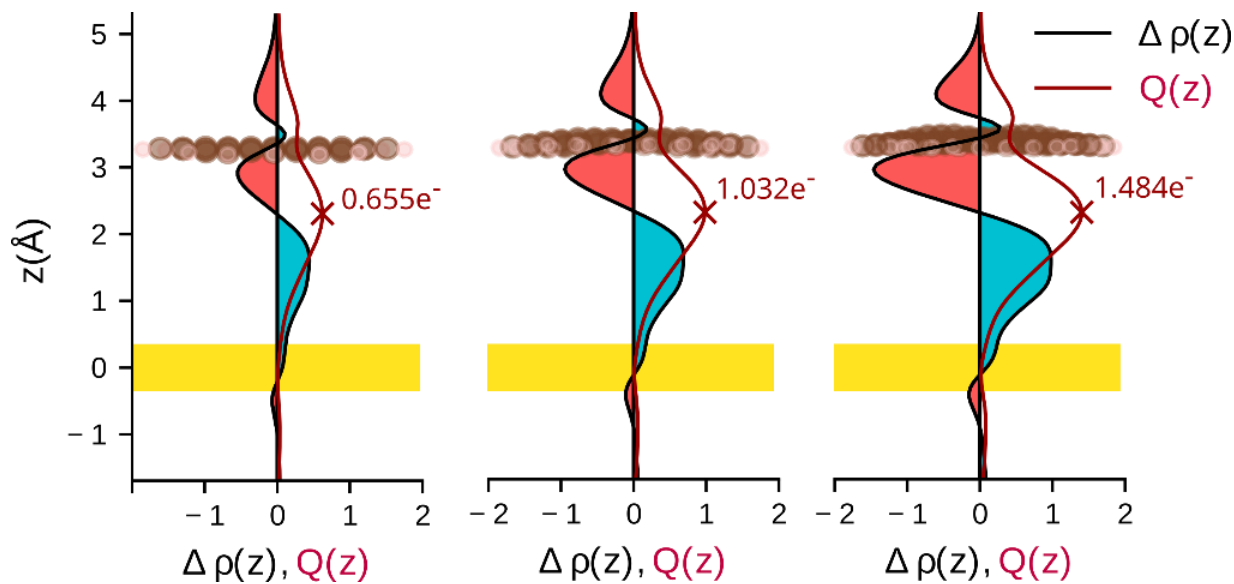

**Figure S12: Plane-averaged charge density differences $\Delta\rho(z)$ and cumulative charge transfer functions $Q(z)$ for the graphene flakes interacting with the Au(111) substrate with the size of 3×3, 4×4 and 5×5.** Areas of electron accumulation are coloured in blue and electron depletion in red. The position of the first substrate layer represented by the yellow rectangles is set to 0. The maximum of $Q(z)$ defines the number of electrons transferred from the flake to the substrate. Brown and white semi-transparent spheres at $z = 3.2$ Å denote z-positions of carbon and hydrogen atoms of the graphene flake, respectively.

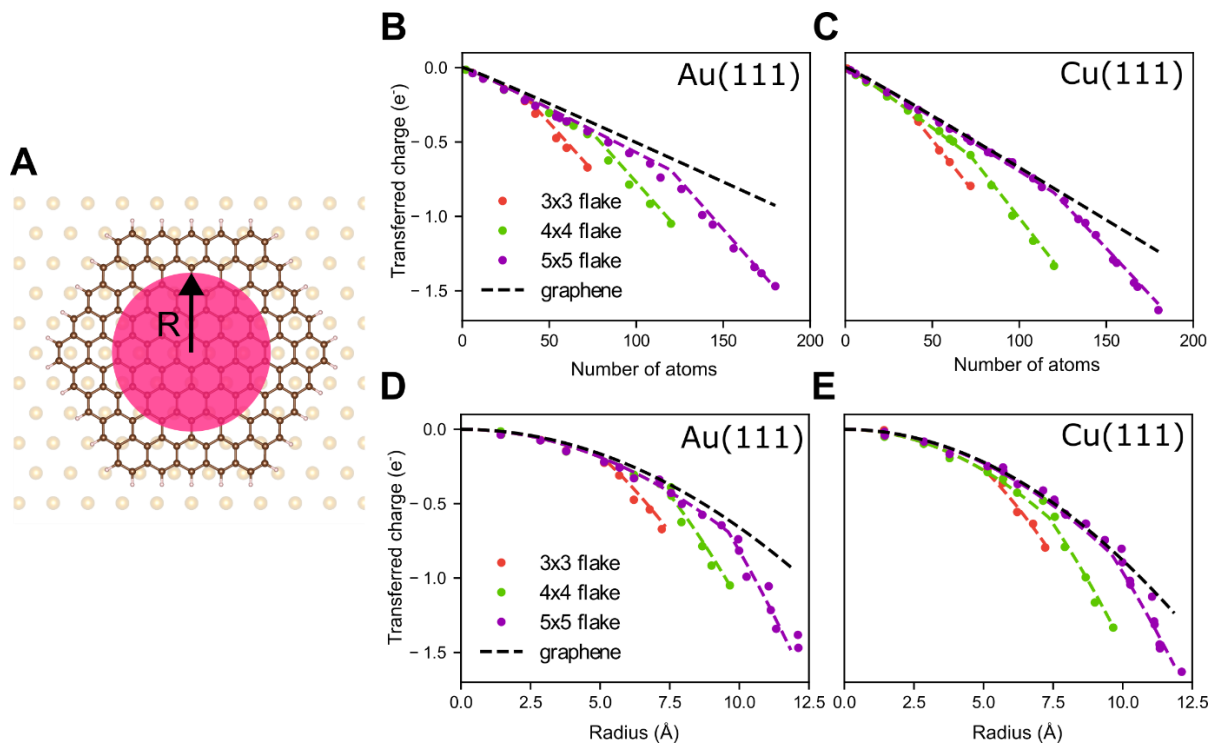

**Figure S13: Analysis of the charge transfer distribution within the graphene flake adsorbed on the Au(111) and Cu(111) substrates.** Panel **A** shows a top view of the graphene flake and the first substrate layer. The pink circle, centred on the graphene flake with a varying radius 'R' defines the area where the transferred charge is evaluated as a function of the number of atoms inside the circle (panels **B**, **C**) and the radius (panels **D**, **E**). Panels **B** and **D** illustrate the transferred charge to the graphene flakes adsorbed on Au(111), while the same for Cu(111) is shown in panels **C** and **E**. Dashed coloured lines in panels **B**-**E** represent fits from our phenomenological model and the dashed black lines show the calculated charge transfer between a graphene and a respective substrate.

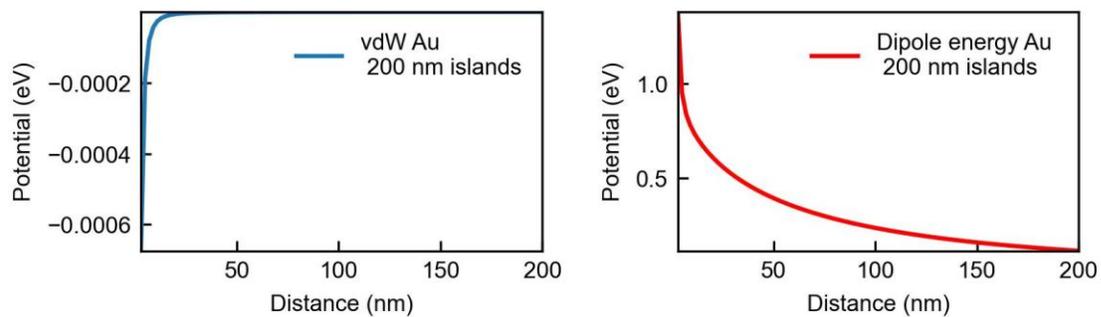

**Fig. S14**: **Calculated potential distribution between 200 nm graphene domains on Au**. A potential-distance dependence resulting from the van der Waals interaction (left) and electrostatic interaction (right) between the domains.

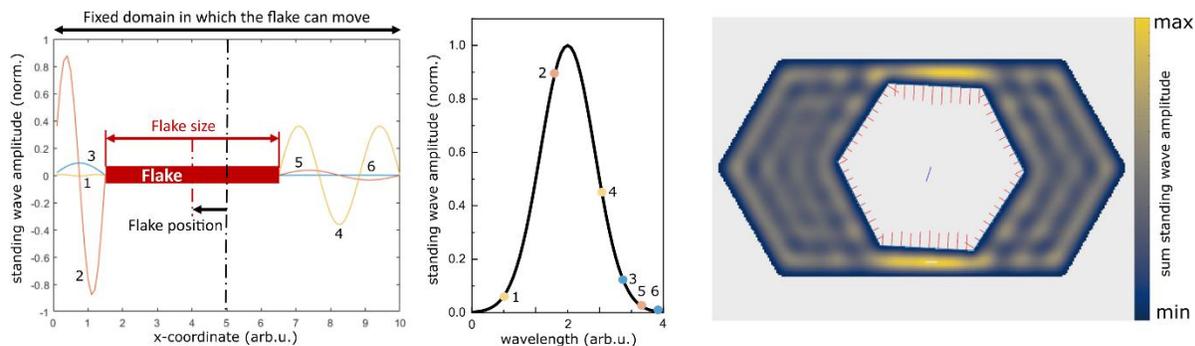

**Fig. S15**: **Explanation of the model of surface undulations**. Left: A 1D sketch illustrating the domain floating on a liquid with undulated surface. The dispersion curve is shown in the middle, with 6 specific wavelengths referring to the waves depicted in the sketch on the left (marked by numbers). Right: The simulations presented in the main text are done in 2D, where we plot the sum intensity of all the surface waves. In this case, due to a confined space in between the compartment edges and the top and bottom edges of the domain, the biggest forces are acting there (red lines), resulting from the waves of highest amplitudes (bright yellow regions).

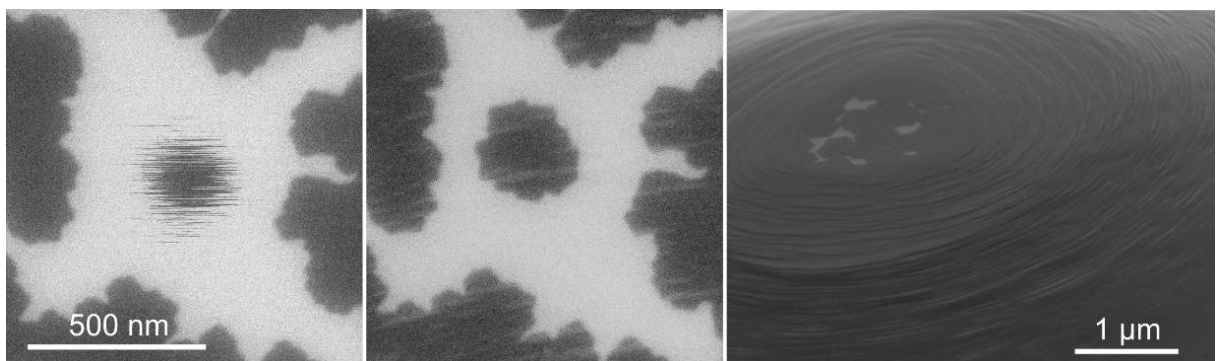

**Figure S16. *In situ* SEM images from a Au sample during the CVD of graphene and after cooling down**. The images reveal surface roughness emergence after solidification. Left: Undercooled liquid gold with graphene domains at 1303 K and 1 Pa ethylene, Middle: the same area just after slow cooling and solidification at 1293 K. Surface exhibits significant roughness, visible as the bright lines across the graphene domains and the substrate. Right: Au droplet that was almost fully covered with graphene in liquid state. As soon as it was cooled to 1273 K and solidified, faceting and roughness emerges (nicely visible in the secondary electron image). This behaviour clearly illustrates the necessity of *in situ* investigations, as the *ex situ* analyses would provide misleading conclusions on e.g. presence of meniscus.

## Description of supplementary movies

**Movies S1: Growth of graphene domians on solid gold.**
*In situ* scanning electron microscopy (SEM) movie acquired at 7 frames/s of graphene domains on solid Au during the deposition of graphene at temperature $T = 1223$ K using ethylene partial pressure $p_{C_2H_4} = 30$ Pa.; The video time $t_{video} = 526$ s.

**Movie S2: Alignment of graphene domain arrays**.
*In situ* scanning electron microscopy (SEM) movie acquired at 7 frames/s of the same graphene domains from Movie S1 on molten Au during the deposition of graphene at temperature $T = 1373$ K using ethylene partial pressure $p_{C_2H_4} = 30$ Pa.; The video time $t_{video} = 188$ s.

**Movie S3: A graphene domain enclosed inside a compartment.** *In situ* SEM movie (7 frames/s) from liquid Au at $T = 1313$ K with $p_{C_2H_4} = 15$ Pa showing wobbling of a graphene domain fully enclosed within a region bounded by other domains. $t_{video} = 23.75$ min.

**Movie S4: Domain dynamics simulation.** A movie showing a growing two-dimensional (2D) hexagonal domain floating on a liquid within a hexagonal compartment (see figure below, left). The dark blue and yellow colors denote low and high amplitudes, respectively, of standing waves formed on the liquid surface. The short red lines indicate forces acting on the domain edges; blue arrow represents magnitude and direction of the total force on the domain. (Center) Plot of the domain position (x,y) and rotation (rot). (Right) Plot of $F_x$ and $F_y$, the $x$- and $y$-components respectively of the total force and the momentum $M_z$ acting on the domain.

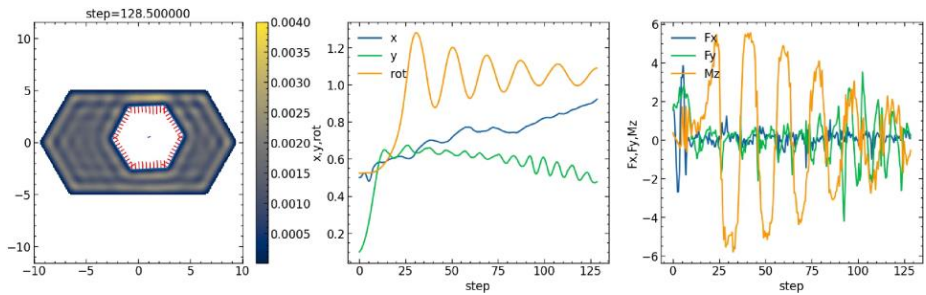

**Movie S5: Domain dynamics simulations with different damping**. Simulated movie showing wobbling of a domain calculated using the same parameters as in Movie S4 but with damping set to one-half of that in Movie S4.

**Movie S6: Rotating graphene domains in an enclosed space.** *In situ* SEM movie (7 frames/s) of molten Cu during graphene growth at $T = 1373$ K with $p_{C_2H_4} = 1 \cdot 10^{-2}$ Pa showing a larger field of view than that in Fig. 3A. $t_{video} = 140$ s.

**Movie S7: Domain assembly inside a graphene compartment on liquid Cu**. *In situ* SEM movie acquired at 7 frames/s during graphene growth on liquid Cu at $T = 1373$ K with $p_{C_2H_4} = 8 \times 10^{-3}$ Pa. $t_{video} = 231$ s.

**Movie S8: Simulation assembly of multiple interacting domains**. A movie showing evolution of six growing domains enclosed within an extended hexagonal space. Initially randomly placed and oriented domains (see figure below, left) spontaneously assemble into a nearly regular array (middle) as a result of surface undulations. With increasing size, the domains attach (right), similar to the experiment shown in Fig. 1B (Movie S2).

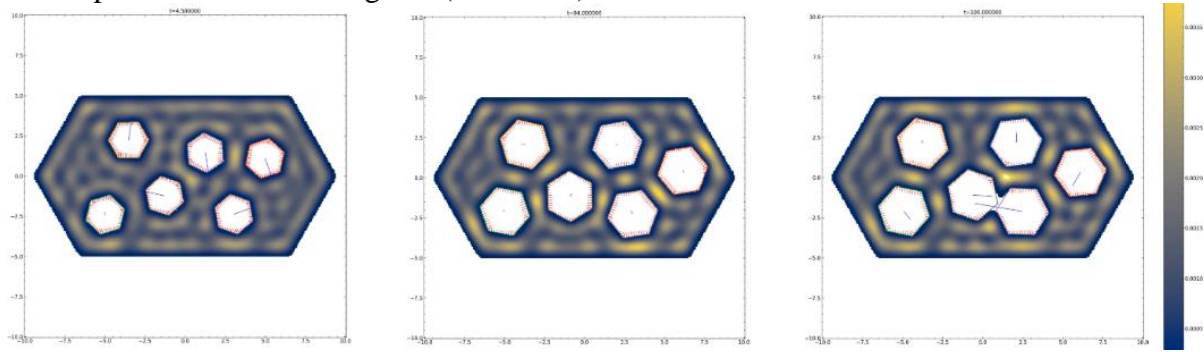